\documentclass[aps,prd,twocolumn,groupedaddress]{revtex4-1}

\usepackage{pstricks,pst-plot,epsfig}
\usepackage{amssymb,enumitem}

\newcommand{\amp}{A_\Phi}

\newcommand{\ben}{\begin{eqnarray}}
\newcommand{\een}{\end{eqnarray}}

\begin{document}


\title{Scalar field as~a~time variable during gravitational evolution}


\author{Anna Nakonieczna}
\email[]{aborkow@kft.umcs.lublin.pl}
\affiliation{Institute of~Physics, Maria Curie-Sk{\l}odowska University, Plac Marii Curie-Sk{\l}odowskiej 1, 20-031 Lublin, Poland \\
Institute of~Agrophysics, Polish Academy of~Sciences, Do{\'s}wiadczalna 4, 20-290 Lublin, Poland}

\author{Jerzy Lewandowski}
\email[]{Jerzy.Lewandowski@fuw.edu.pl}
\affiliation{Faculty of Physics, University of Warsaw, Pasteura 5, 02-093 Warszawa, Poland}


\date{\today}

\begin{abstract}
Using a~scalar field as~an~intrinsic 'clock' while investigating the~dynamics of~gravitational systems has been successfully pursued in~various researches on~the~border between classical and~quantum gravity. The~objective of~our~research was to~check explicitly whether the~scalar field can serve as~a~time variable during dynamical evolution of~the~matter-geometry system, especially in~regions of~high curvature, which are essential from the~perspective of~quantum gravity. For~this purpose, we~analyzed a~gravitational collapse of~a~self-interacting scalar field within the~framework of~general relativity. The~obtained results indicated that the~hypersurfaces of~constant scalar field are spacelike in~dynamical regions nearby the~singularities formed during the~investigated process. The~scalar field values change monotonically in~the~areas, in~which the~constancy hypersurfaces are spacelike.
\end{abstract}


\maketitle


\section{Introduction}
\label{sec:Intro}

Canonical gravity is an~approach to~general relativity based on~an arbitrary slicing of~spacetime into spacelike hypersurfaces labeled by a~single parameter, which can be interpreted as~a~time variable. All possible slicings are physically equivalent. The~dynamical evolution in~such a~system is governed by the~Hamiltonian, which is also a~constraint and~a~generator of~gauge transformations between the~spacelike slices. The~quantizations of~gravity based on~its canonical formulation encounter the~so-called problem of~time. It is related to~the~fact that there is no clear notion of~time in~gravitational systems, which could be transferred from the~classical to~quantum level. The~problem reveals itself as~crucial in~attempts to~define the~dynamics of~a~quantized system.

The~idea of~using matter to~quantify time during the~course of~gravitational evolution was proposed by DeWitt in~\cite{DeWitt1967}. In~the~context of~quantum theory of~gravity, it~was suggested that the~matter content of~a~spacetime can play a~role of~a~'clock' when investigating the~dynamical behavior of~geometry. One of~the~degrees of~freedom of~the~system is interpreted as~a~dynamical observer. It serves as~a~reference for~the~remaining degrees of~freedom, whose dynamics is followed with respect to~it. Potentially, this concept can be also applied inversely, i.e.,~the~geometry could be used to~introduce a~physical time when the~dynamical evolution of~matter is examined.

The~notion of~internal time is especially important in~non-perturbative quantum gravity, which deals with a~problem of~investigating time evolution without fixed geometry of~a~background spacetime~\cite{Rovelli1991}. A~Hamiltonian which describes quantum evolution of~a~gravitational field with respect to~time measured by a~'physical clock' provided by spacelike constancy hypersurfaces of~a~scalar field was presented in~\cite{RovelliSmolin1994}. It was defined through a~regularization procedure based on~the~Ashtekar variables~\cite{Ashtekar1986,Ashtekar1987} and~the~loop representation~\cite{RovelliSmolin1988,RovelliSmolin1990} and~was shown to~be finite and~diffeomorphism invariant. Recently, the~idea was developed by applying loop quantum gravity tools. The~scalar field, which is an~intrinsic element of~the~model, was treated as~the~time variable for~the~relational Dirac observables, whose dynamics could be traced with respect to~the~field~\cite{DomagalaGieselKaminskiLewandowski2010,LewandowskiDomagalaDziendzikowski2012}.

The~scalar field was also used to~quantify the~passage of~time during quantum cosmological evolutions. The~studies of~the~inflation epoch of~the~Universe in~the~regime of~loop quantum gravity was studied with the~use of~a~scalar field with an~arbitrary potential~\cite{AlexanderMaleckiSmolin2004}. The~gauge for~the~Hamiltonian constraints was chosen so~that the~scalar field was constant on~spacelike hypersurfaces. Treating the~scalar field as~a~time variable allowed writing the~Hamiltonian constraints in~the~form of~a~time-dependent Schr{\"o}dinger equation. A~similar approach and~a~massless scalar field is used in~the~loop quantum cosmology~\cite{AshtekarPawlowskiSingh2006PRL,AshtekarPawlowskiSingh2006PRD}.

The~Brans-Dicke scalar field, which is non-minimally coupled to~geometry, was also examined as~a~'clock' in~the~quantum cosmological context~\cite{ZhangArtymowskiMa2013}. It was claimed that since the~field is a~monotonic function of~cosmological time, it~may serve as~an~internal time variable. A~spatially flat and~isotropic cosmological Brans-Dicke model was quantized within the~framework of~loop quantum cosmology and~the~effective Hamiltonian was thus obtained.

A~homogeneous, massless, minimally coupled scalar field was used to~calculate the~tunneling decay rate~\cite{DabrowskiLarsen1995} of~a~simple harmonic universe~\cite{GrahamHornKachruRajendranTorroba2014}. It~is a~cosmological model, which oscillates due to~the~presence of~a~negative cosmological constant and~a~matter component with repulsive gravity. An~additional field with negligible contribution to~the~total energy density of~the~system, which served as~a~'clock', was included within it~\cite{MithaniVilenkin2012,MithaniVilenkin2015}. Its presence was necessary, because the~scale factor, which is the~only dynamical variable of~the~model, is not monotonic in~an oscillating universe and~hence it~cannot be treated as~a~time variable.

The~version of~the~Wheeler-DeWitt equation, which uses the~scalar field as~a~variable common for~internal time and~a~Hamiltonian evolution parameter, and~its quantum counterpart for~the~timelike case were presented in~\cite{Perlov2015}. The~former describes the~evolution of~four-dimensional Lorentz hypersurfaces embedded in~a~five-dimensional spacetime, along the~massless scalar field. The~latter has the~form of~the~Schr{\"o}dinger equation, i.e.,~it contains the~internal time derivative on~the~right-hand side and~presents the~evolution with respect to~it.

Apart from the~scalar field, dust and~radiative fluid were also used to~provide a~matter degree of~freedom, which allowed tracing the~temporal evolution of~a~gravitational system. The~comoving coordinates of~dust particles and~the~proper time along the~dust world lines served as~canonical coordinates in~the~phase space of~the~dynamical system~\cite{BrownKuchar1995}. The~issue of~unitarity of~the~evolution proceeding in~the~isotropic and~homogeneous Brans-Dicke quantum cosmological model, with a~time variable defined by the~matter fluid, was discussed in~\cite{AlmeidaBatistaFabrisMoniz2015}.

Despite the~important achievements in~formulating a~consistent description of~a~time evolution of~a~gravitational system, which were enabled by treating matter as~a~time variable, a~question about the~relevance of~such an~approach remains open. In~general, it~is assumed that the~behavior of~matter is desirable during the~investigated part of~an evolution, i.e.,~that the~spacelike character of~the~chosen spacetime slices is preserved and~that the~selected time parametrization remains monotonic during its course. However, the~arguments in~favor of~these assumptions are limited to~certain cases (such as~homogeneity of~a~scalar field during an~initial phase of~inflation, which results in~a~spacelike character of~constant field hypersurfaces~\cite{AlexanderMaleckiSmolin2004}) and~there does not exist a~general justification for~them. While for~perturbations of~the~cosmological spatially homogeneous solutions it~is plausible that the~constancy surfaces of~the~scalar field are still spacelike, a~character of~those surfaces is highly non-trivial in~the~case of~the~spherically symmetric gravitational collapse. The~latter is also a~hot
subject of~quantum gravity~\cite{GambiniPullin2015}.

In~the~present paper we considered an~evolution of~a~scalar field, which was its dynamical gravitational collapse, within the~framework of~a~four-dimensional formulation of~general relativity. The~main objective of~the~research was to~examine the~feasibility of~using the~scalar field as~a~time variable in~spacetime regions of~high curvature, which are essential for~the~applications of~the~quantum theory of~gravity. The~problem was posed and~its theoretical formulation was introduced in~Section~\ref{sec:TheorFrame}. Section~\ref{sec:NumComp} contains the~description of~numerical algorithm used during computations, the~specifics of~its implementation and~its tests. The~results were presented and~discussed in~Section~\ref{sec:Results}, while conclusions were gathered in~Section~\ref{sec:Conclusions}.

\section{Theoretical framework}
\label{sec:TheorFrame}

The~Lagrangian of~the~scalar field $\Phi$ is
\ben
\mathcal{L} = \nabla_\alpha\Phi \nabla_\beta\Phi g^{\alpha\beta}
\een
and its stress-energy tensor has the~form
\ben
T_{\mu\nu} = \nabla_\mu\Phi \nabla_\nu\Phi - \frac{1}{2} g_{\mu\nu} g^{\alpha\beta} \nabla_\alpha\Phi \nabla_\beta\Phi,
\een
where $g_{\mu\nu}$ is a~metric tensor of~spacetime. The~equation of~motion for~the~scalar field derived from the~variational principle is
\ben
\Box\Phi = 0.
\label{eqn:sf-general}
\een

The~gravitational collapse was investigated in~spherical symmetry. Since the~evolving field is massless, it~is convenient to~follow the~process in~double null coordinates $x^\mu=\left(u,v,\theta,\phi\right)$, in~which the~line element has the~form
\ben
ds^2 = -a^2 dudv + r^2 d\Omega^2,
\een
where $a$ and~$r$ are arbitrary functions of~the~retarded time $u$ and~the~advanced time $v$, while $d\Omega^2=d\theta^2 + \sin^2\theta d\phi^2$ is the~line element of~the~unit sphere, $\theta$ and~$\phi$ are angular coordinates. The~employed method is referred to~as~the~2+2~formalism~\cite{Winicour2009}.

The~scalar field equation (\ref{eqn:sf-general}) written in~the~selected coordinates is
\ben
r\Phi_{,uv} + r_{,u}\Phi_{,v} + r_{,v}\Phi_{,u} = 0
\label{eqn:sf-dn}
\een
and the~Einstein equations for~the~gravitational field are the~following:
\ben
rr_{,uv} + r_{,u}r_{,v} + \frac{a^2}{4} = 0, \\
\frac{a_{,uv}}{a} - \frac{a_{,u}a_{,v}}{a^2} + \frac{r_{,uv}}{r} + \Phi_{,u}\Phi_{,v} = 0, \\
r_{,uu} - 2\frac{a_{,u}r_{,u}}{a} + r\Phi_{,u}^2 = 0, \\
r_{,vv} - 2\frac{a_{,v}r_{,v}}{a} + r\Phi_{,v}^2 = 0,
\label{eqn:gf-dn}
\een
where $_{,x^\mu}$ denotes the~partial derivative.

The~system of~equations (\ref{eqn:sf-dn}) -- (\ref{eqn:gf-dn}) describes the~spherically symmetric dynamical gravitational collapse of~a~scalar field in~double null coordinates. The~particulars of~the~numerical algorithm employed to~solve the~above equations and~its tests are presented in~Section~\ref{sec:NumComp}.

\begin{figure*}
\includegraphics[width=0.315\textwidth]{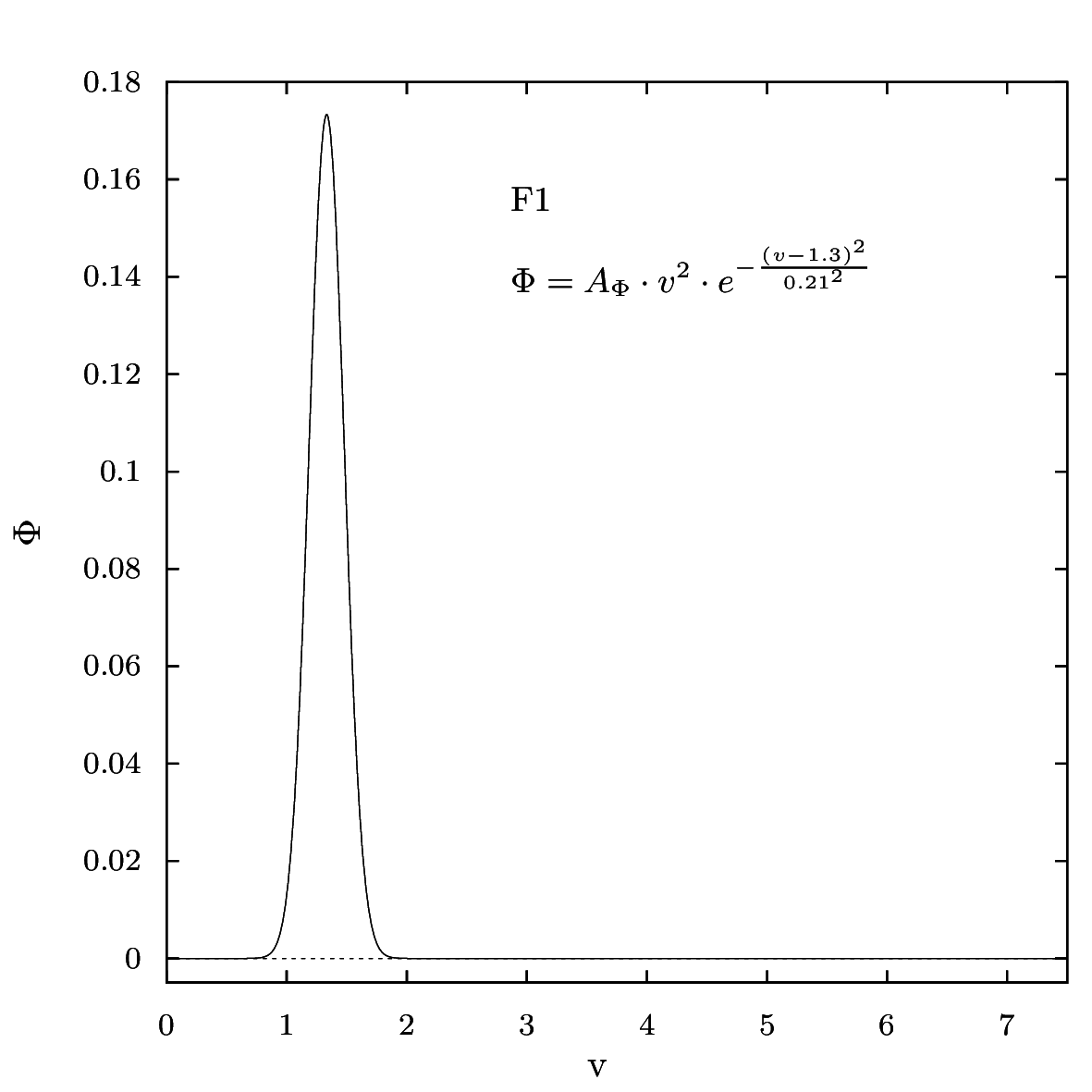}
\includegraphics[width=0.315\textwidth]{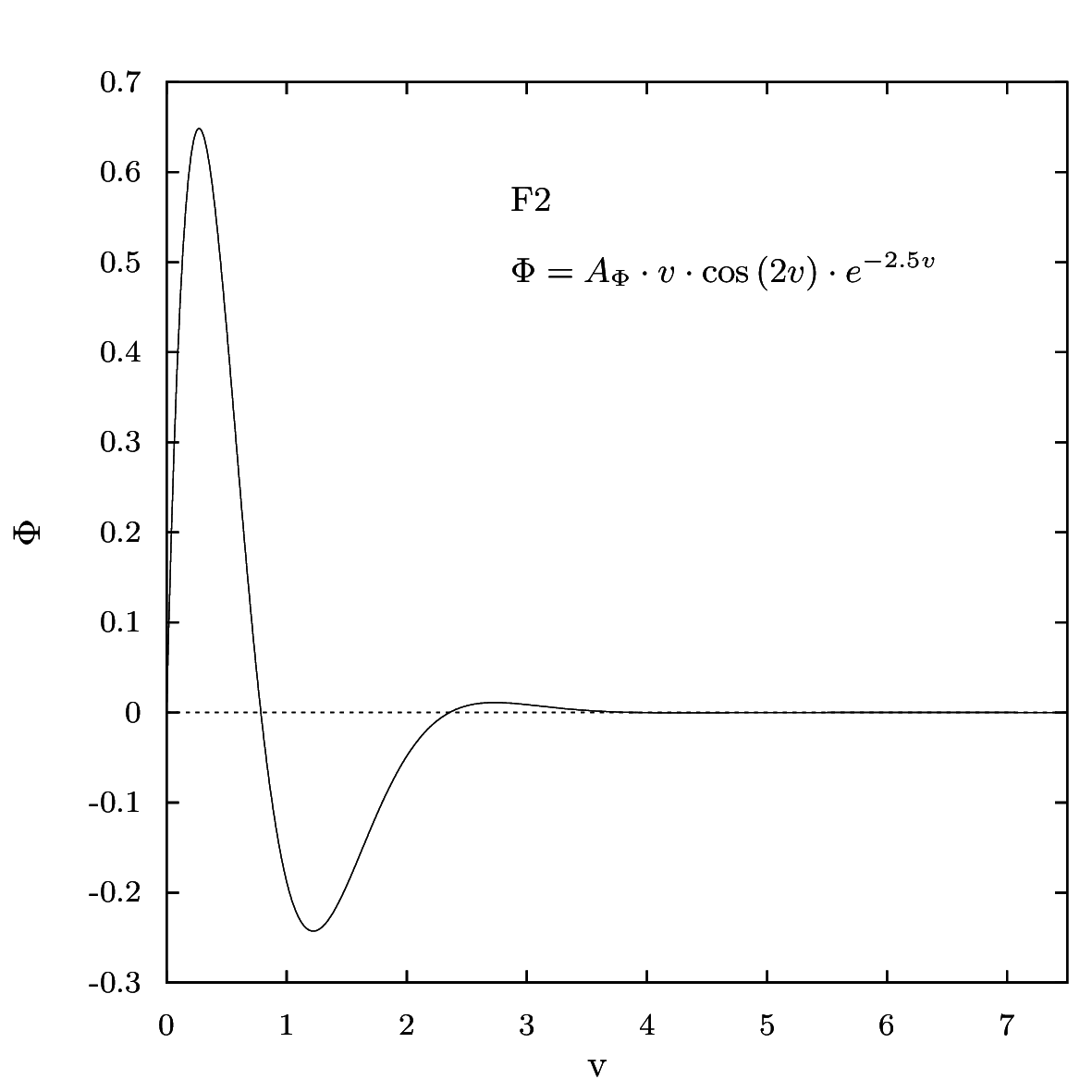}
\includegraphics[width=0.315\textwidth]{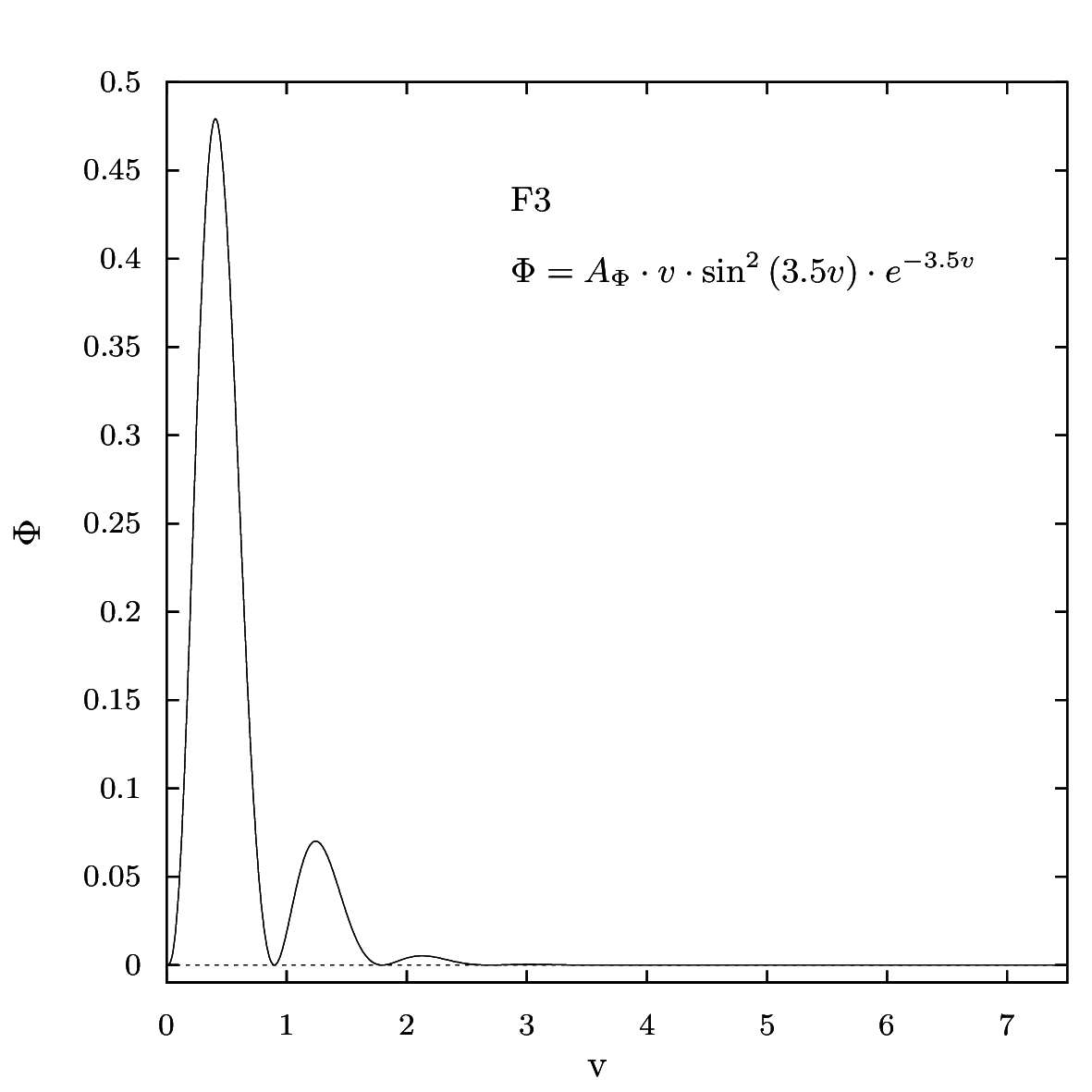}
\includegraphics[width=0.315\textwidth]{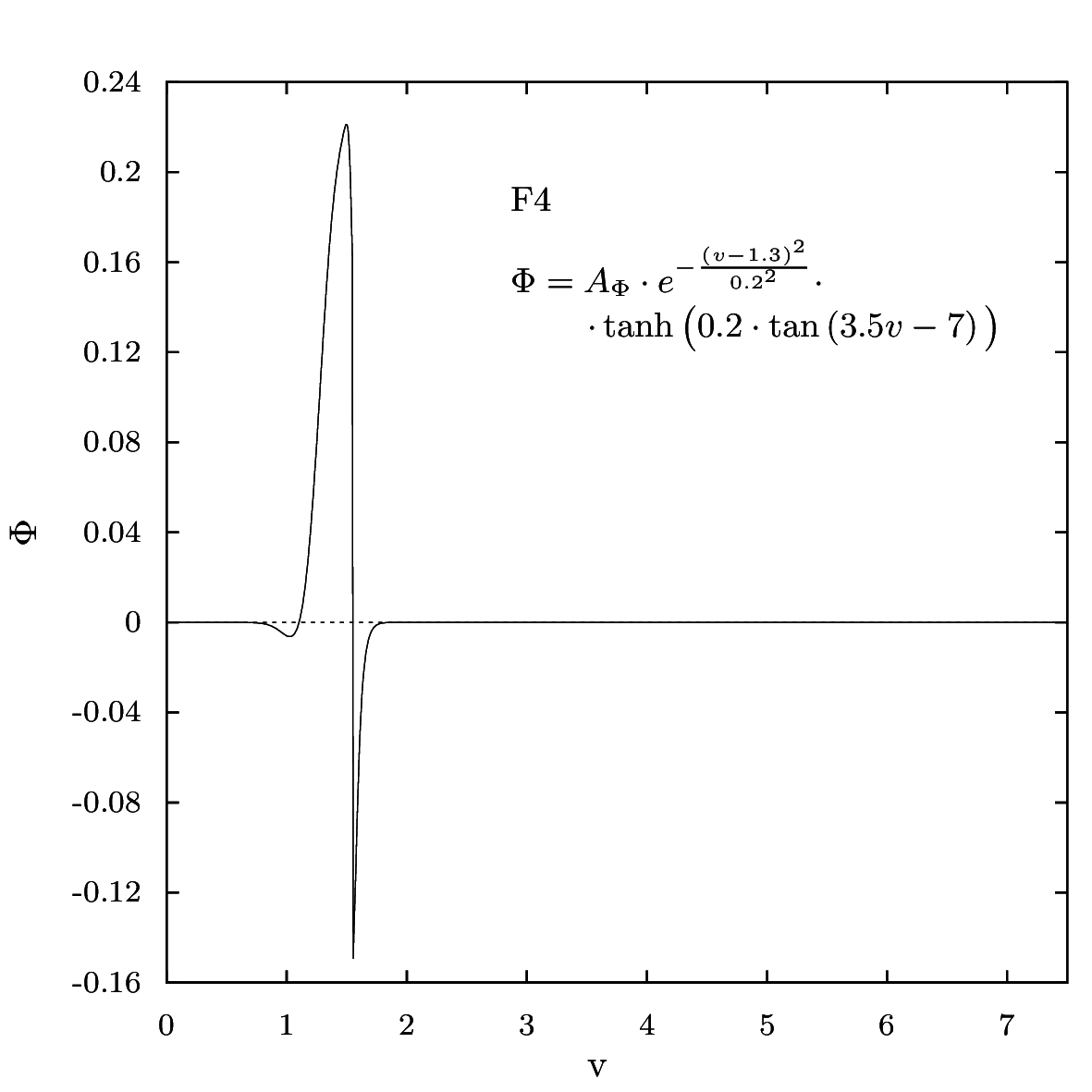}
\includegraphics[width=0.315\textwidth]{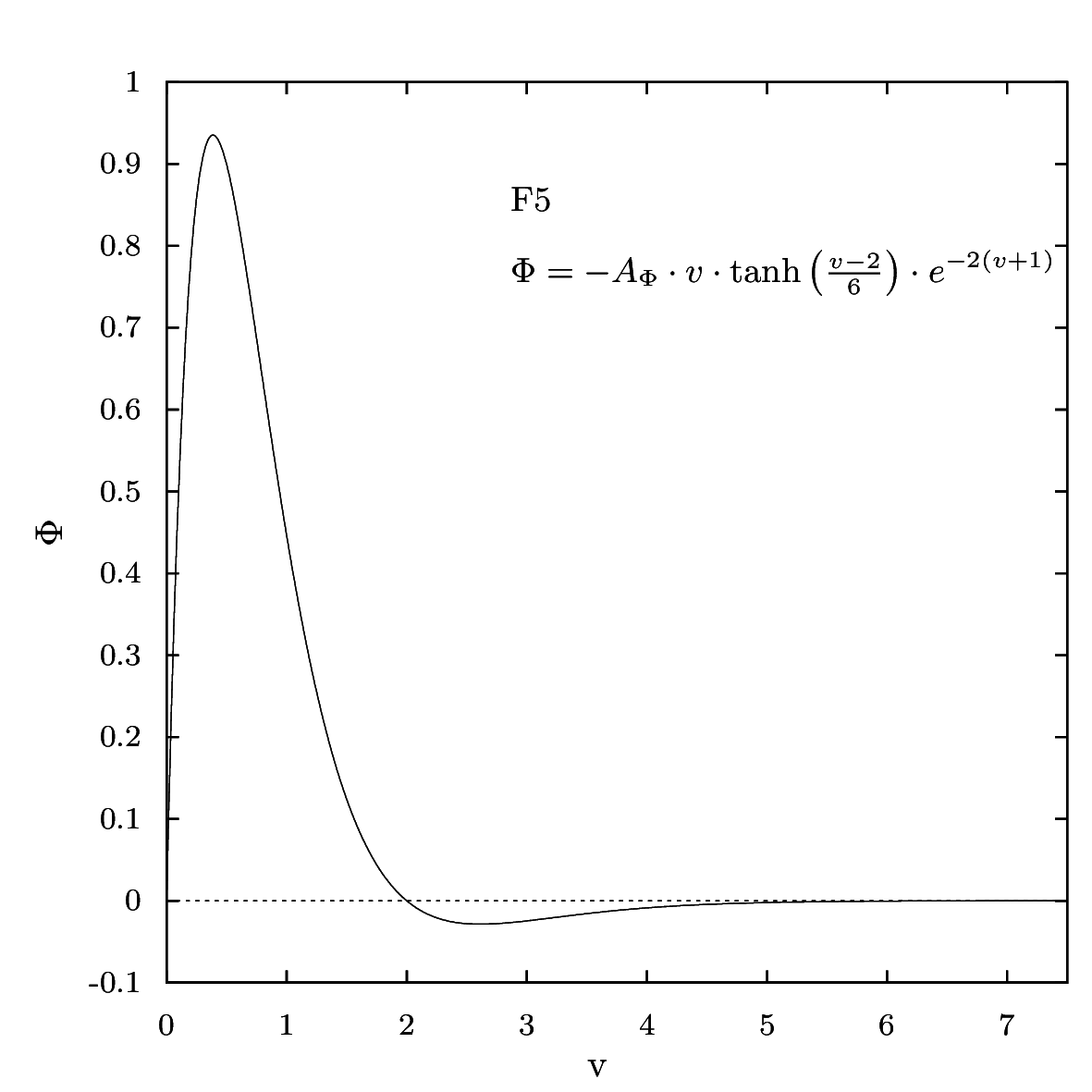}
\includegraphics[width=0.315\textwidth]{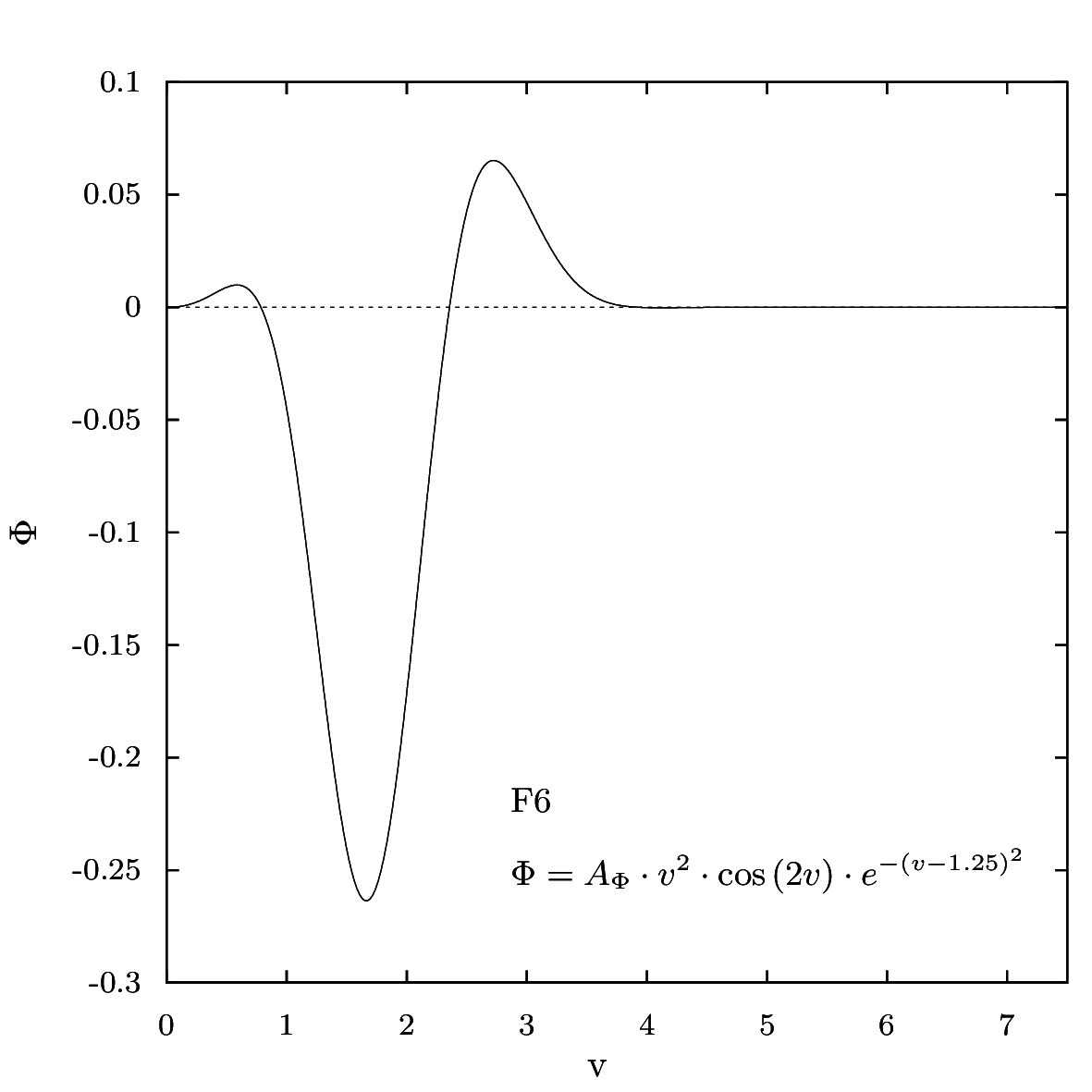}
\includegraphics[width=0.315\textwidth]{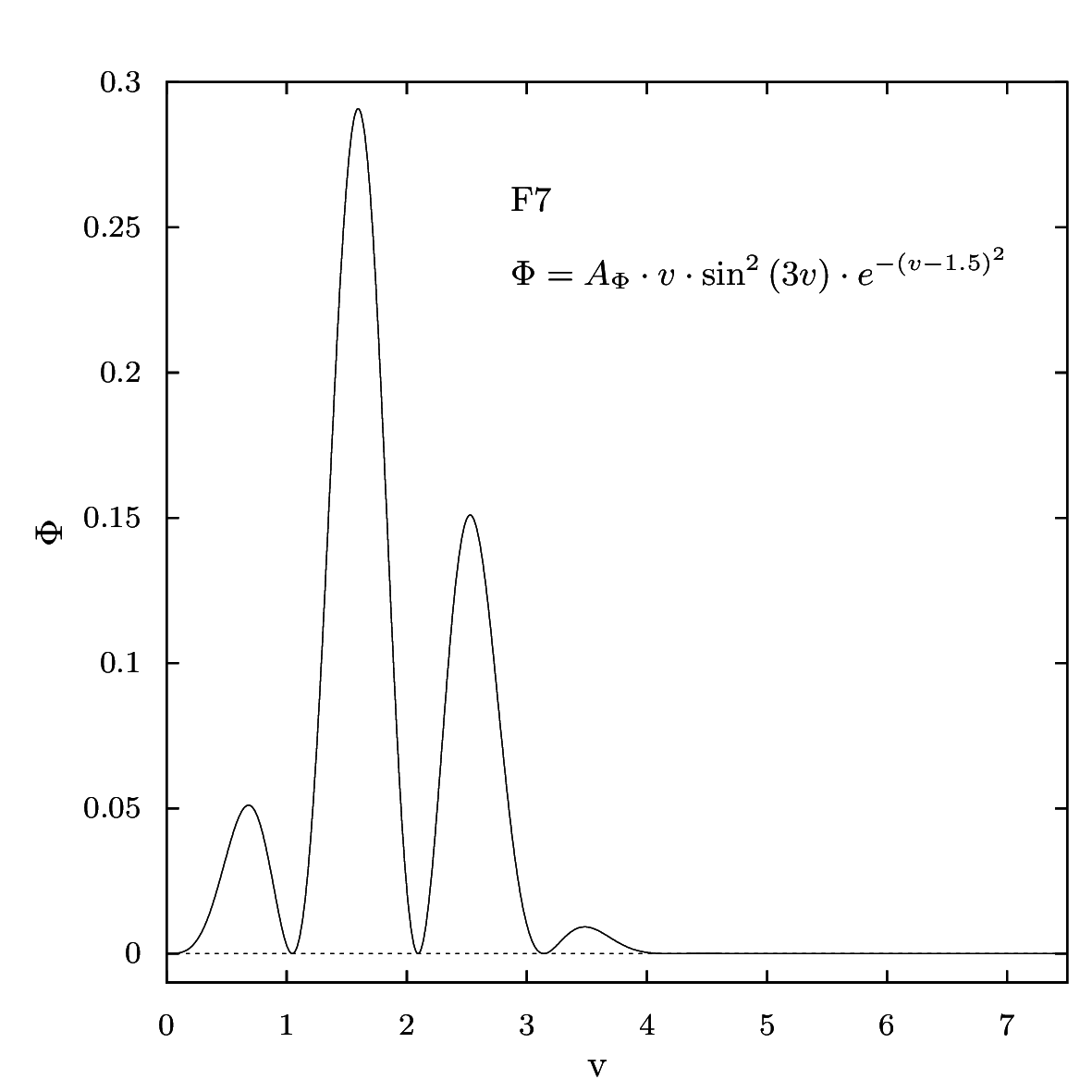}
\includegraphics[width=0.315\textwidth]{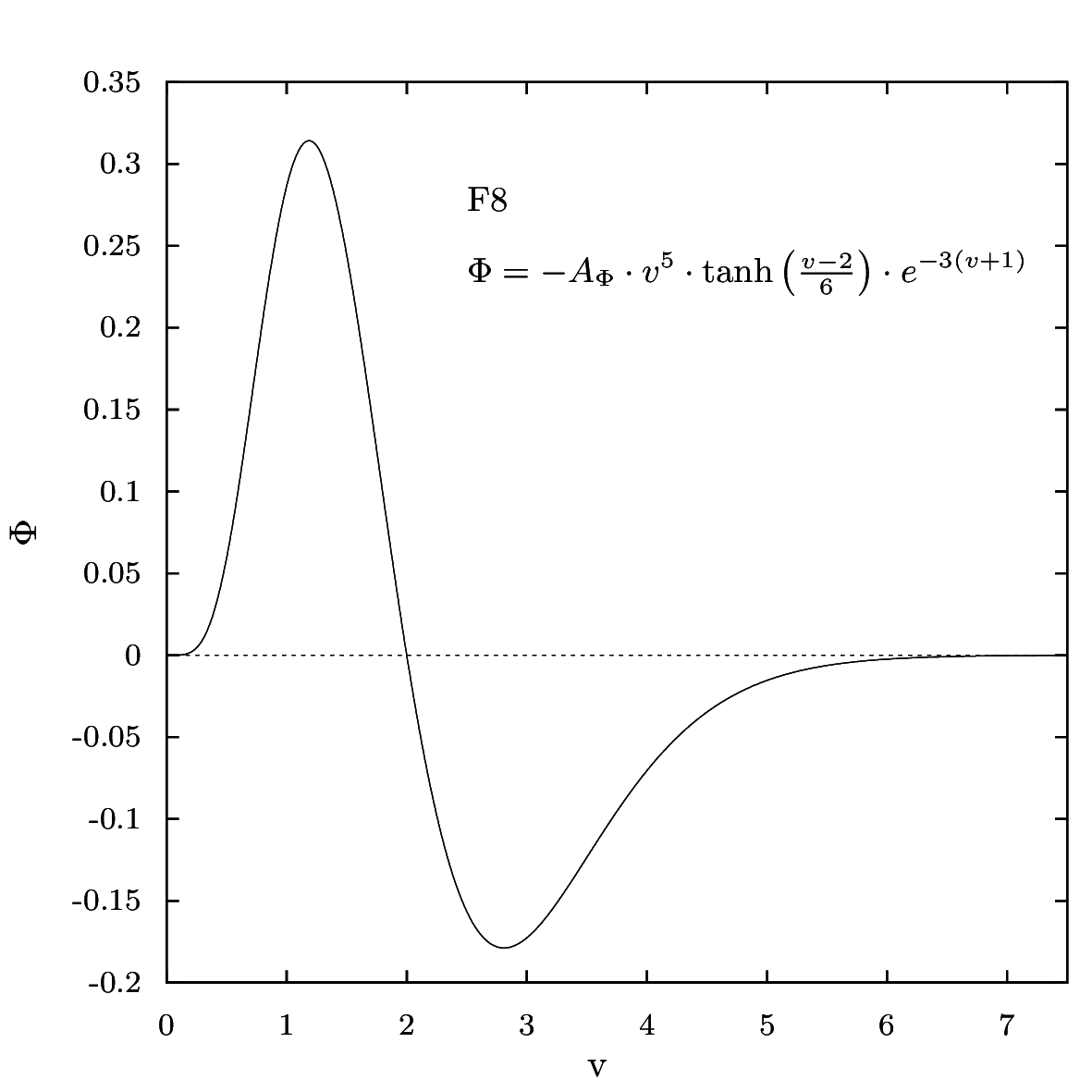}
\includegraphics[width=0.315\textwidth]{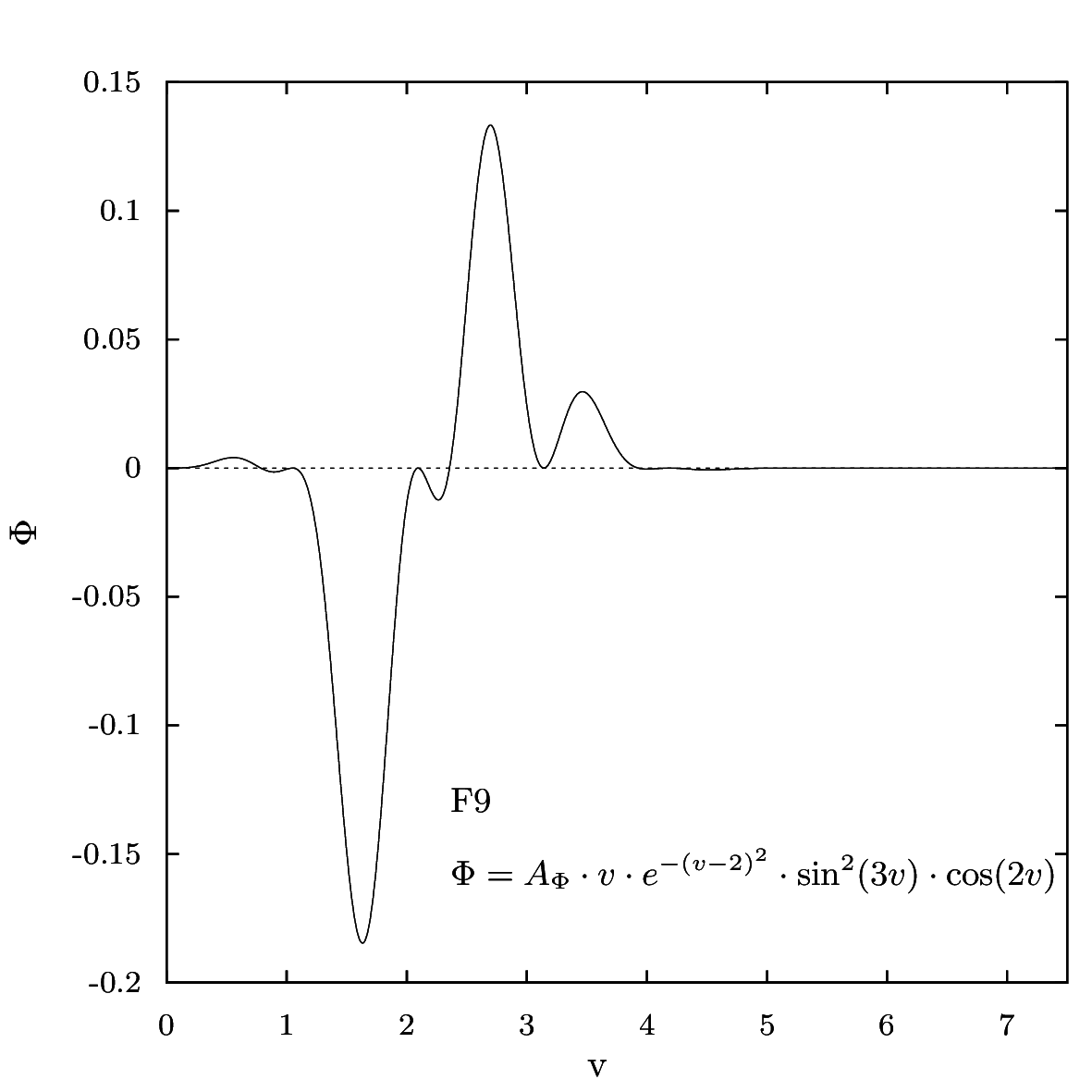}
\caption{Families of~the~scalar field initial profiles $\Phi\left(0,v\right)$. The~amplitudes $\amp$ were set as~equal to~$0.1$, $5.5$, $5.0$, $0.75$, $147.5$, $0.115$, $0.185$, $700$ and~$0.135$ for~families F1 to~F9, respectively.}
\label{fig:InitialProfiles}
\end{figure*}

The~only arbitrary initial data of~the~computations is the~profile of~the~evolving scalar field, which is imposed on~the~null $u=const.$ hypersurface. The~initial hypersurface will be denoted as~$u=0$ for~further convenience. The~investigated families of~initial profiles are presented in~Fig.~\ref{fig:InitialProfiles}. They are one-parameter classes of~functions parametrized by the~amplitude $\amp$. The~values of~$\amp$ used in~the~plots were chosen so~that the~collapse resulted in~a~singular spacetime, as~will be explained in~more detail below. From now on, the~profiles presented in~Fig.~\ref{fig:InitialProfiles} with the~amplitudes specified in~its caption will be referred to~as profiles P1 to~P9. The~profiles describe spherically symmetric shells of~matter as~they tend to~zero for~$v\rightarrow\infty$ and~$v\rightarrow 0$. The~former condition ensures asymptotic flatness of~spacetime. The~latter is equivalent to~vanishing as~$r\rightarrow 0$, because the~gauge freedom of~equations in~$v$-coordinate was thus fixed, see Section~\ref{sec:NumComp}.

\section{Numerical computations}
\label{sec:NumComp}

The~considered dynamical gravitational collapse is described by second-order differential equations (\ref{eqn:sf-dn}) -- (\ref{eqn:gf-dn}). In~order to~solve the~system numerically a~set of~auxiliary variables was introduced
\ben
c = \frac{a_{,u}}{a}, \quad d = \frac{a_{,v}}{a}, \quad f = r_{,u}, \quad g = r_{,v}, \nonumber\\ x = \Phi_{,u}, \quad y = \Phi_{,v}.
\label{eqn:substitution}
\een
The~quantities $f$ and~$g$ are related to~the~expansions $\theta_u$ and~$\theta_v$ of~the~null geodesic vector fields $\partial_u$ and,~respectively, $\partial_v$ via $\theta_u=2r^{-1}f$ and~$\theta_v=2r^{-1}g$~\cite{MaedaHaradaCarr2009}. The~hypersurfaces determined by~$f=0$ and~$g=0$ are the~trapping and~anti-trapping horizons in~the~spacetime, respectively~\cite{Hayward1994}. The~squares of~the~quantities $x$ and~$y$ represent fluxes of~the~scalar field through null hypersurfaces of~constant $v$ and~$u$, respectively. The~substitutions (\ref{eqn:substitution}) allowed us to~obtain the~system of~first-order differential equations, which determines the~course of~the~investigated process
\ben
\label{eqn:first}
& V1: & a_{,u} - ac = 0,\\
& V2: & a_{,v} - ad = 0,\\
& V3: & r_{,u} - f = 0,\\
& V4: & r_{,v} - g = 0,\\
& V5: & \Phi_{,u} - x = 0,\\
& V6: & \Phi_{,v} - y = 0,\\
& E1: & f_{,u} = 2cf - rx^2,\\
& E2: & g_{,v} = 2dg - ry^2,\\
& E3: & f_{,v} = g_{,u} = -\lambda r^{-1},\\
& E4: & c_{,v} = d_{,u} = \lambda r^{-2} - xy,\\
& S: & rx_{,v} = ry_{,u} = -\eta,
\label{eqn:last}
\een
where $\lambda=\frac{a^2}{4}+fg$ and~$\eta=gx+fy$. The~evolution of~$d$ and~$y$ along $u$ was governed by relations $E4$ and~$S$, respectively. The~remaining quantities, that is $a$, $\Phi$, $g$, $r$, $f$ and~$x$, evolved along the~$v$-coordinate according to~equations $V2$, $V6$, $E2$, $V4$, $E3$ and~$S$. The~function $c$ did not play an~active role during the~computations, so~it~was ignored. 

The~evolution was followed on~a~two-dimensional grid constructed in~the~$\left(vu\right)$-plane with integration steps in~$v$ and~$u$ directions denoted as~$h_v$ and~$h_u$, respectively, and~initially equal to~$10^{-4}$. The~numerical algorithm proposed in~\cite{HamadeStewart1996} was implemented. The~general form of~equations (\ref{eqn:first}) -- (\ref{eqn:last}) can be written as
\ben
\textsf{f}_{,u} = \textsf{F}\left(\textsf{f},\textsf{g}\right), \qquad
\textsf{g}_{,v} = \textsf{G}\left(\textsf{f},\textsf{g}\right).
\een
Calculating a~value of~the~particular function $\textsf{f}\left(u,v\right)$ or $\textsf{g}\left(u,v\right)$ at~the~point $\left(v_0,u_0\right)$ required the~knowledge of~the~values of~functions \textsf{F} and~\textsf{G} at~points $\left(v_0-h_v,u_0\right)$ and~$\left(v_0,u_0-h_u\right)$
\ben
\textsf{f}\big\arrowvert_{\left(v_0,u_0\right)} = \frac{1}{2}\bigg(\textsf{ff}\big\arrowvert_{\left(v_0,u_0\right)} +
\textsf{f}\big\arrowvert_{\left(v_0,u_0-h_u\right)} + \nonumber\\
+ h_u\textsf{F}\left(\textsf{ff},\textsf{gg}\right)\big\arrowvert_{\left(v_0,u_0\right)}\bigg),
\label{eqn:algorithm-first}\\
\textsf{g}\big\arrowvert_{\left(v_0,u_0\right)} = \frac{1}{2}\bigg(\textsf{gg}\big\arrowvert_{\left(v_0,u_0\right)} +
\textsf{g}\big\arrowvert_{\left(v_0-h_v,u_0\right)} + \nonumber\\
+ h_v\textsf{G}\left(\textsf{ff},\textsf{gg}\right)\big\arrowvert_{\left(v_0,u_0\right)}\bigg),
\een
where two types of~auxiliary quantities were introduced
\ben
\textsf{ff}\big\arrowvert_{\left(v_0,u_0\right)} = \textsf{f}\big\arrowvert_{\left(v_0,u_0-h_u\right)} + \nonumber\\
+ h_u\textsf{F}\left(\textsf{f},\textsf{g}\right)\big\arrowvert_{\left(v_0,u_0-h_u\right)}, \\
\textsf{gg}\big\arrowvert_{\left(v_0,u_0\right)} = \textsf{g}\big\arrowvert_{\left(v_0-h_v,u_0\right)} + \nonumber\\
+ \frac{h_v}{2}\bigg(\textsf{G}\left(\textsf{f},\textsf{g}\right)\big\arrowvert_{\left(v_0,u_0\right)} +
\textsf{G}\left(\textsf{ff},\textsf{gg}\right)\big\arrowvert_{\left(v_0,u_0\right)}\bigg).
\label{eqn:algorithm-last}
\een

Double null coordinates ensured regular behavior of~all the~evolving quantities within the~domain of~integration except the~vicinity of~$r=0$. During numerical computations considerable difficulties also arose nearby the~event horizon, where the~function $f$ was discontinuous. A~dense numerical grid was necessary to~conduct the~calculations in~the~neighborhood of~the~event horizon and~to~examine the~behavior of~fields beyond it, especially for~large values of~advanced time. Since regions in~which the~dense grid was required were well-defined within the~domain of~integration, an~adaptive grid was implemented in~order to~perform calculations with a~smaller step where necessary. An algorithm which enabled us to~make the~grid denser in~the~$u$-direction was implemented, as~it~was sufficient for~the~conducted analyses~\cite{BorkowskaRogatkoModerski2011}. A~local truncation error indicator was the~function $\Delta_u r\cdot r^{-1}$, where $\Delta_u r$ is a~difference between the~values of~$r$ for~the~fixed $v$ at~two adjacent hypersurfaces of~constant $u$. It~was used to~mark areas where the~denser mesh was required because it~changed its value significantly in~adequate regions~\cite{OrenPiran2003}.

Boundary conditions of~the~process were determined along the~line $u=v$. The~spacetime containing a~spherically symmetric shell of~matter is flat inside the~shell and~at~large radii from it. This fact justified an~assumption that the~line $u=v$ was not affected by the~presence of~collapsing matter and~its vicinity was nearly flat, which gave $r=0$ along $u=v$. Using the~equation $E3$ the~boundary condition $\lambda=0$ was obtained, which in~combination with equations $V3$ and~$V4$ led to~$f=-g=-\frac{a}{2}$. The~requirement $\Phi_{,r}=a_{,r}=0$ along $u=v$ guaranteed the~flattening of~the~functions nearby $r=0$, which made the~numerical analysis possible. Since $r$ was a~nonlinear function of~both null coordinates, the~three-point regressive derivative method with a~variable step was used to~implement this boundary condition, apart from the~first point where the~Euler's method was used. Due to~the~form of~the~equation~$S$ the~relation $\eta=0$ held along $u=v$. The~definition of~$\eta$ gave $x=y$ along this line. Boundary conditions for~the~quantities evolving along the~$u$-coordinate were obtained according to~the~equation (\ref{eqn:algorithm-first}) with auxiliary quantities (\ref{eqn:algorithm-last}) taken as~equal to~the~corresponding functions at~$r=0$.

Initial conditions were formulated along an~arbitrarily chosen null hypersurface, denoted as~$u=0$. The~assumption of~a~negligible spacetime curvature in~the~area where the~collapse begins justified the~condition $d\left(0,v\right)=0$, which finally fixed the~remaining gauge freedom in~$v$-coordinate. The~flatness of~spacetime in~the~vicinity of~the~line $u=v$, which was discussed above, gave $a\left(0,v\right)=1$. As~was mentioned in~Section~\ref{sec:TheorFrame}, the~only arbitrary initial condition was the~profile of~the~scalar field $\Phi\left(0,v\right)$. The~quantity $y\left(0,v\right)$ was computed analytically using the~equation $V6$. The~evolution of~functions $g$, $r$, $f$ and~$x$ was followed using the~three-point Simpson's method except the~first point, where the~Newton's method was implemented.

The~accuracy of~the~program was tested in~two ways, namely the~convergence of~the~obtained results and~mass conservation in~spacetime were checked. In~order to~monitor the~convergence of~the~code computations for~all the~examined initial profiles (presented in~Fig.~\ref{fig:InitialProfiles}) were conducted on~four grids, whose integration steps were multiples of~$10^{-4}$. A~step of~a~particular grid was twice the~size of~a~denser one. The~convergence was examined on~$u=const.$ hypersurfaces chosen for~each profile arbitrarily. The~values of~retarded time were selected so~that the~hypersurface was situated close to~the~event horizon, but in~the~region where the~adaptive grid was yet inactive, which was necessary for~performing a~proper comparison of~results.

\begin{figure}
\includegraphics[width=0.3835\textwidth]{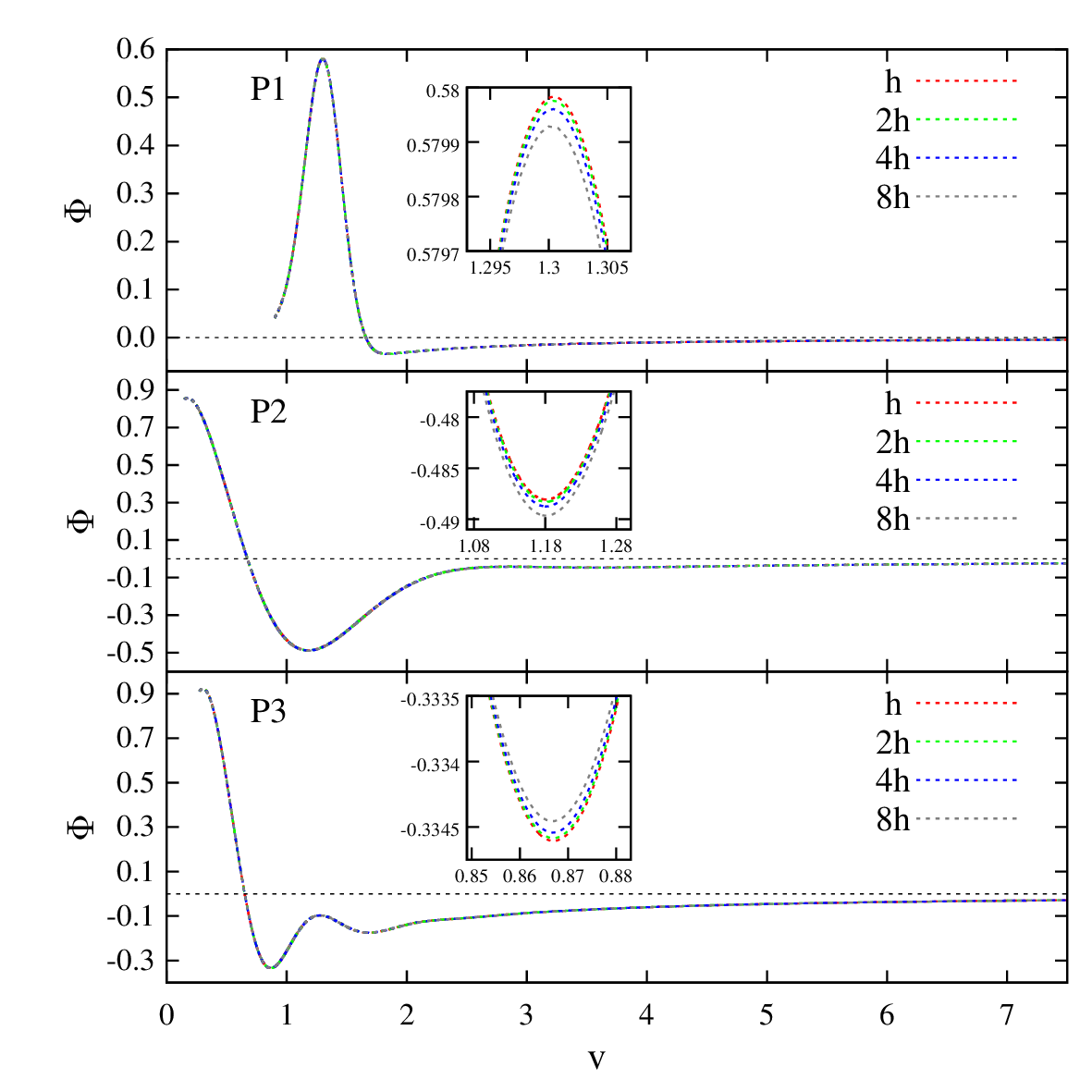}
\includegraphics[width=0.3835\textwidth]{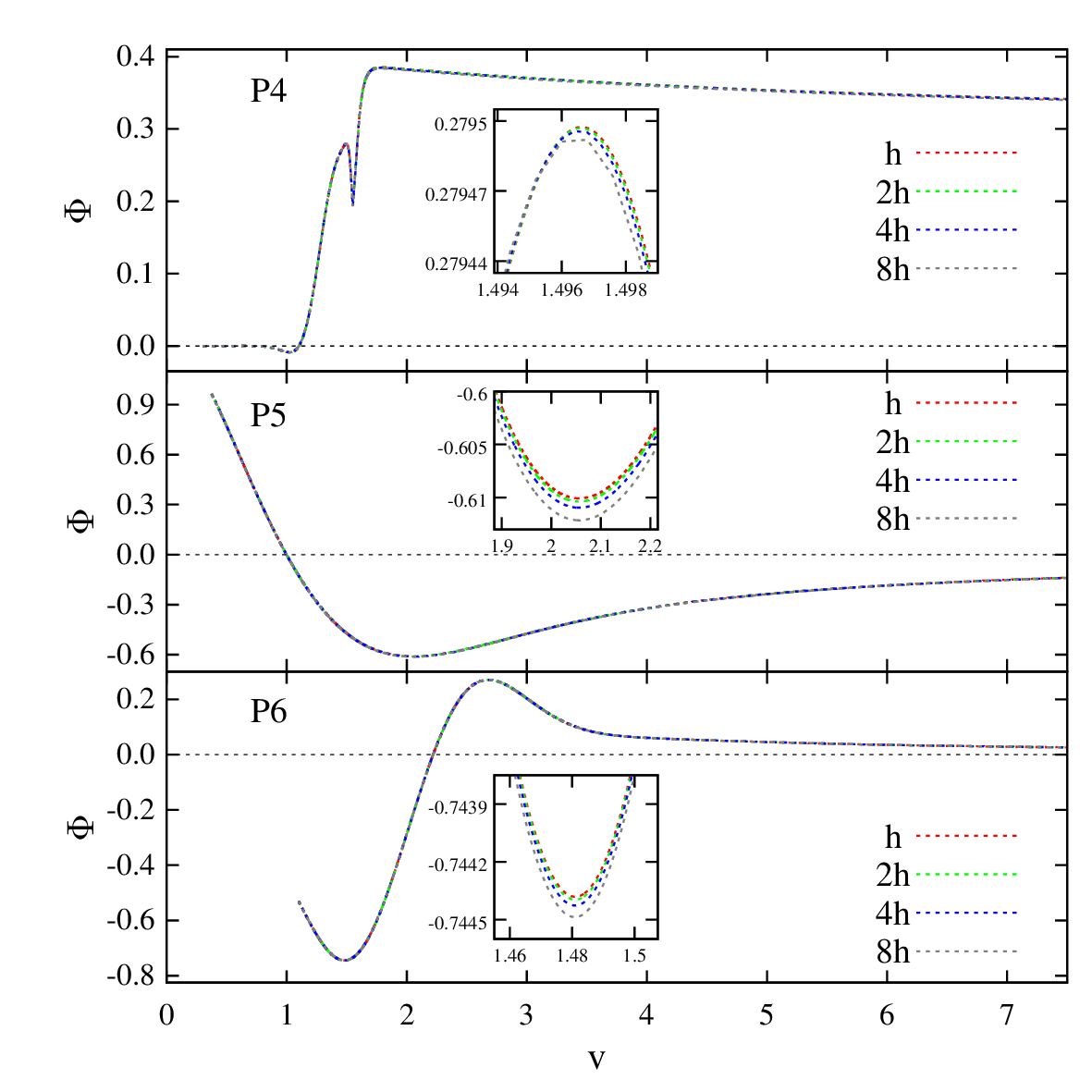}
\includegraphics[width=0.3835\textwidth]{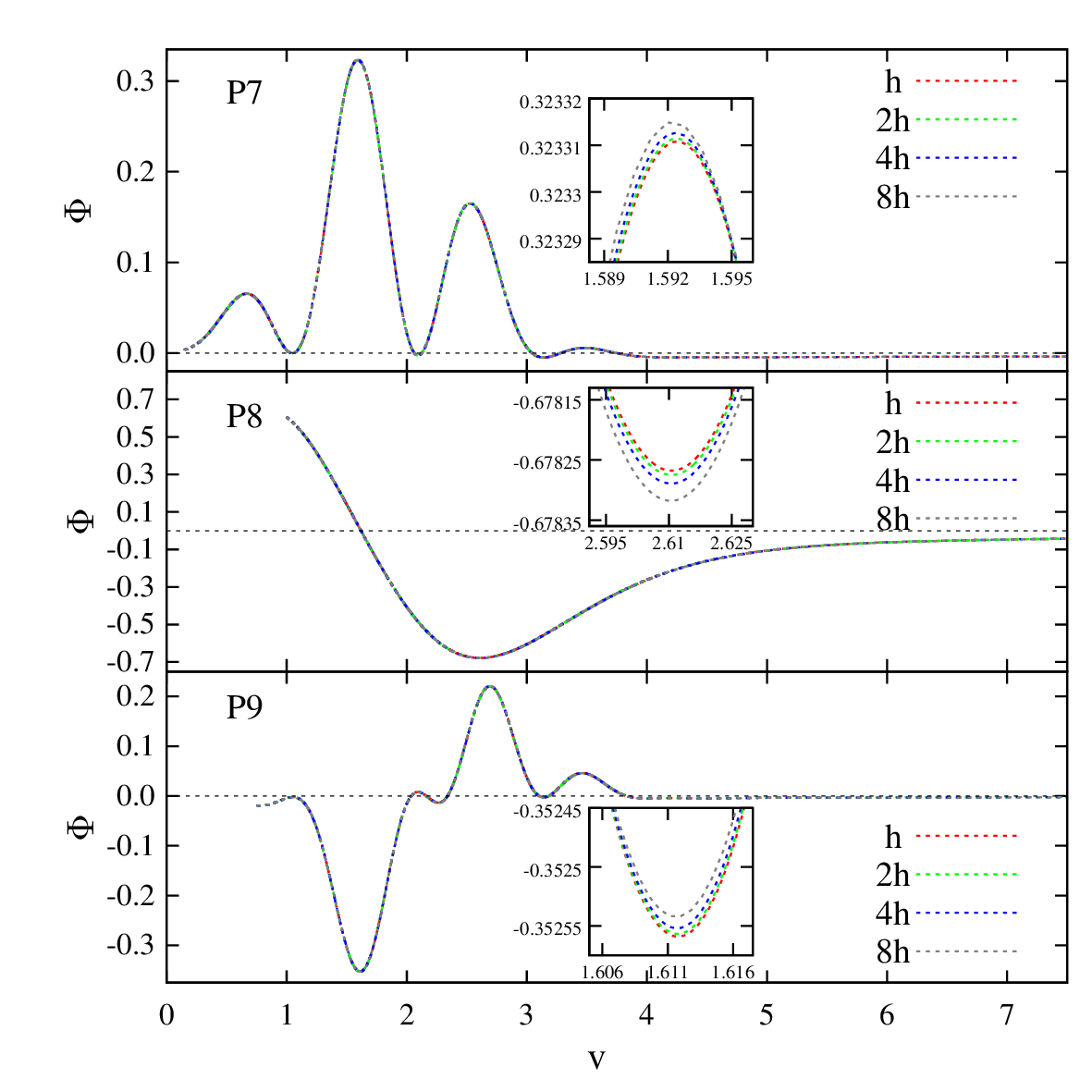}
\caption{(Color online) The~convergence of~the~scalar field function during the~collapse initiated with the~examined profiles. The~functions $\Phi\left(v\right)$ were plotted for~evolutions conducted with integration steps, which were multiples of~\mbox{$h=10^{-4}$}, along hypersurfaces of~constant $u$ equal to~$0.9$, $0.1504$, $0.2752$, $0.3$, $0.3752$, $1.1$, $0.1504$, $1.0$ and~$0.7504$ for~profiles~P1 to~P9, respectively.}
\label{fig:Conv1}
\end{figure}

\begin{figure}
\includegraphics[width=0.395\textwidth]{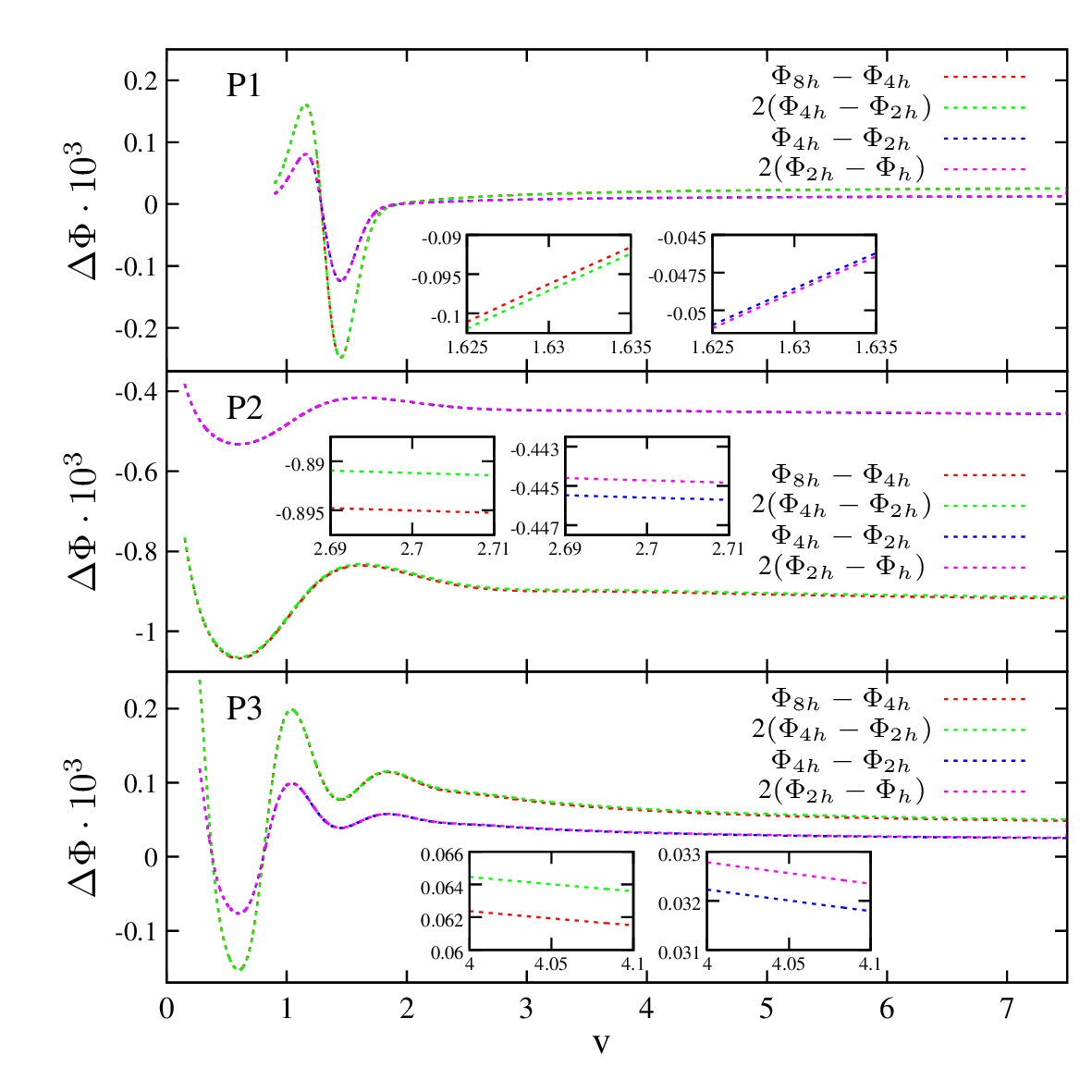}
\includegraphics[width=0.395\textwidth]{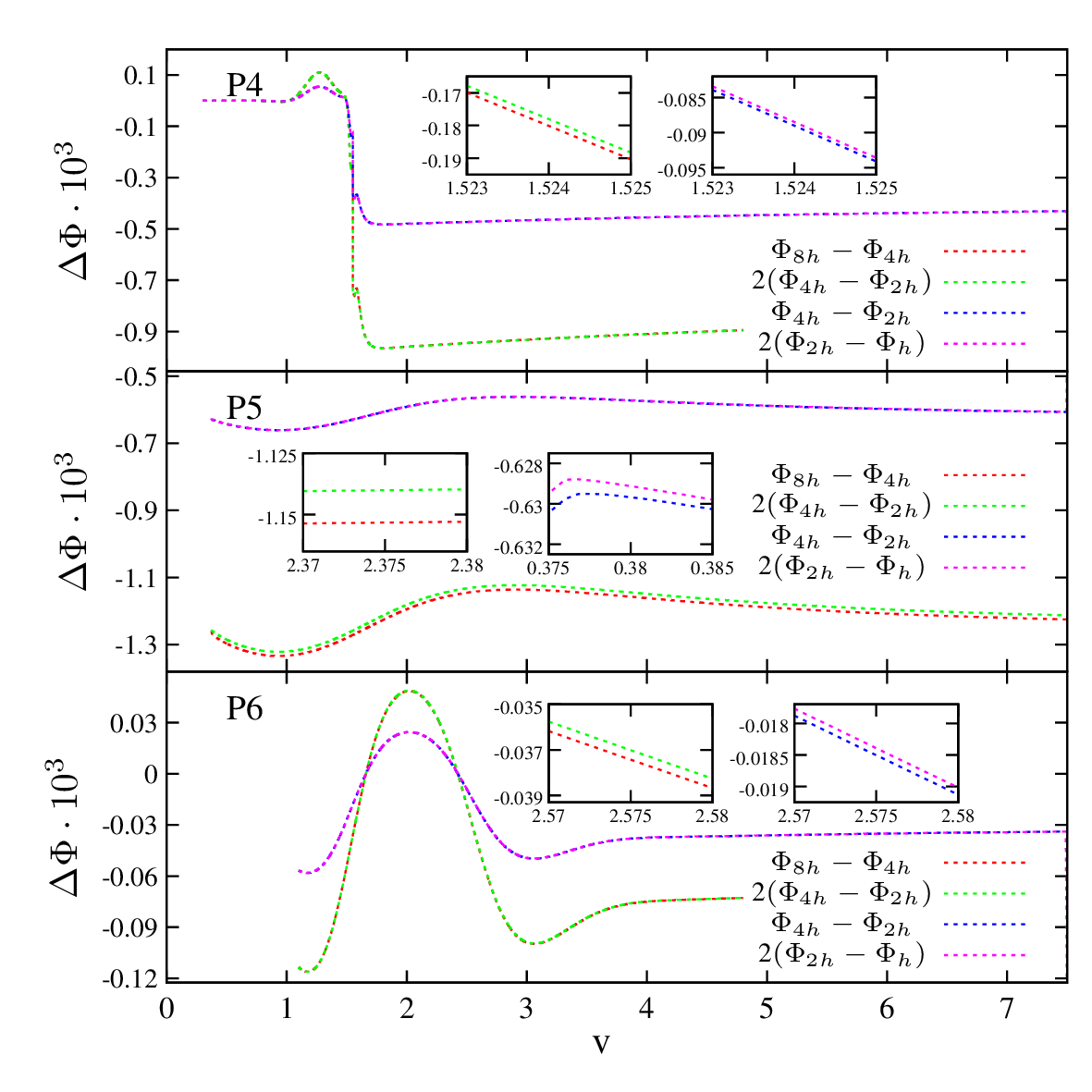}
\includegraphics[width=0.395\textwidth]{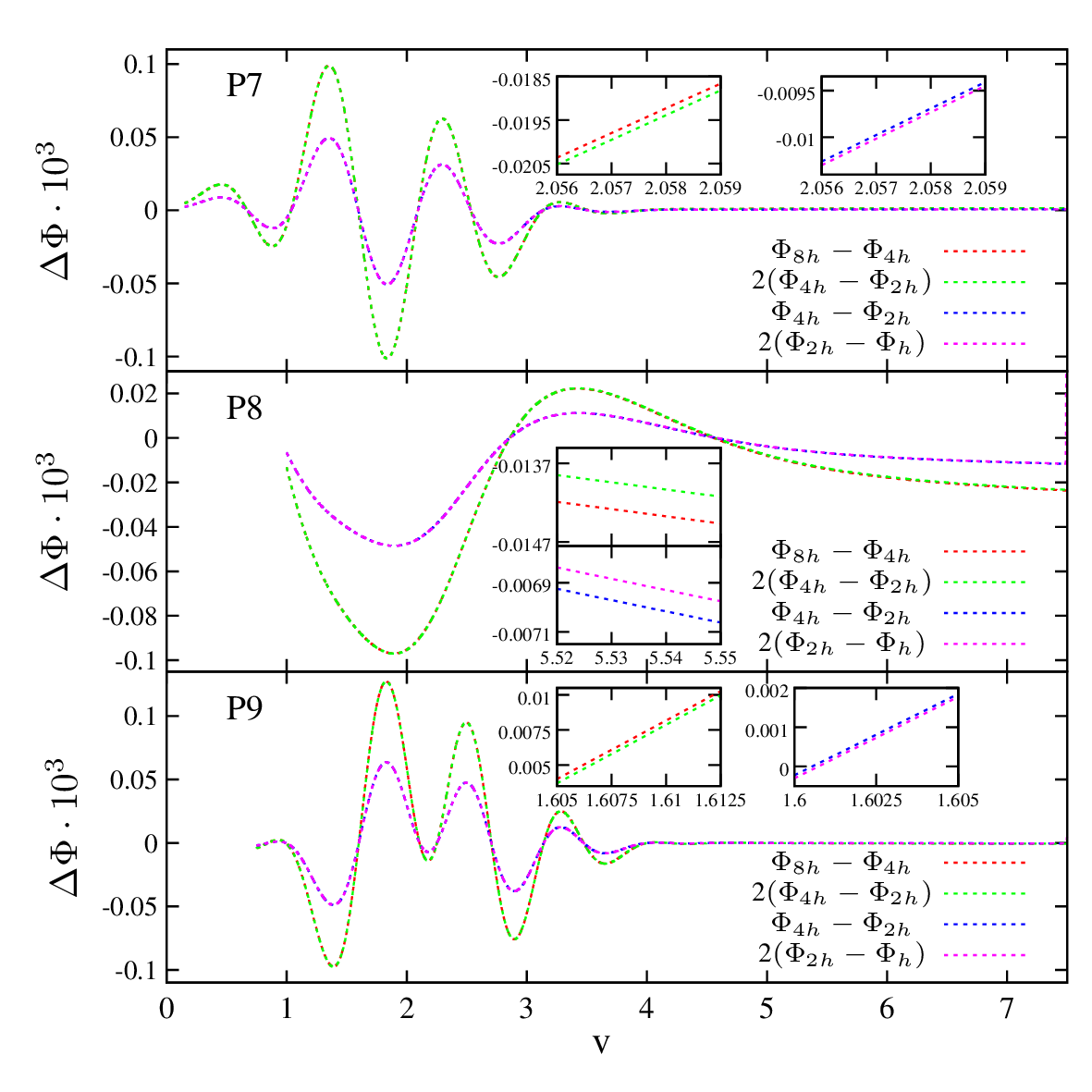}
\caption{(Color online) The~linear convergence of~the~code. The~differences between field functions calculated on~grids with different integration steps (multiples of~\mbox{$h=10^{-4}$}) and~their doubles were obtained for~profiles~P1 to~P9 along the~same hypersurfaces of~constant $u$ as~in~Fig.~\ref{fig:Conv1}.}
\label{fig:Conv2}
\end{figure}

The~field functions along the~selected hypersurfaces of~constant~$u$ are shown in~Fig.~\ref{fig:Conv1}. The~areas in~which the~differences among functions obtained on~various grids were most significant were magnified. The~maximum discrepancy between the~finest and~coarsest grids equal to~$2.6\%$ was observed for~the~profile P5. The~remaining profiles displayed even better behavior, because the~above-mentioned differences did not exceed~$1.3\%$. Figure~\ref{fig:Conv2} presents the~linear convergence of~the~numerical code. The~maximal divergence between profiles obtained on~two grids with a~quotient of~integration steps equal to~$2$ and~their respective doubles was~$3.4\%$. As~expected, the~errors became smaller linearly as~the~grid density increased. The~convergence analysis revealed that both the~proposed algorithm and~the~numerical code were adequate for~solving the~system of~equations (\ref{eqn:first}) -- (\ref{eqn:last}) with all investigated initial profiles~P1 to~P9.

\begin{figure}
\includegraphics[width=0.365\textwidth]{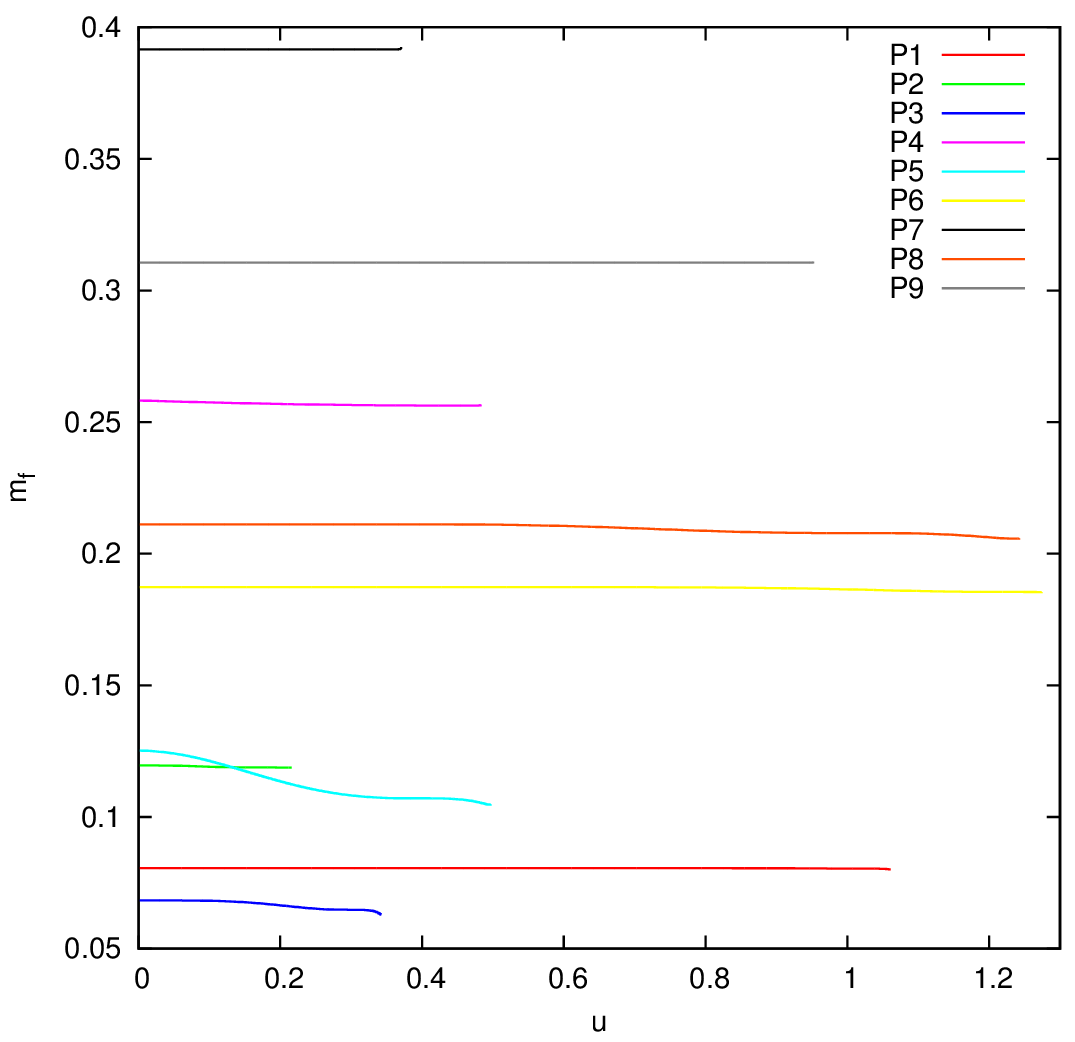}
\caption{(Color online) Hawking masses calculated along $v=7.5$, $m_f$, as~functions of~retarded time, $u$, for~spacetimes obtained with initial profiles~P1 to~P9.}
\label{fig:MassCons}
\end{figure}

The~mass conservation in~the~obtained spacetimes was investigated with the~use of~a~notion of~Hawking mass~\cite{Hawking1968}. It~is a~quasi-local mass function, which for~the~Schwarzschild-type spacetimes in~double null coordinates can be written as~\cite{HamadeStewart1996}
\ben
m\left(u,v\right) = \frac{r}{2} \left( 1 + \frac{4r_{,u}r_{,v}}{a^2} \right) = \frac{r}{2} \left( 1 + \frac{4}{a^2}fg \right)
\label{eqn:HawkingMass}
\een
and~it~describes the~mass included in~a~sphere of~the~radius $r\left(u,v\right)$. The~Hawking mass as~a~function of~retarded time along the~line $v=7.5$, which was a~maximal value of~advanced time achieved numerically, for~all the~investigated families of~initial profiles is presented in~Fig.~\ref{fig:MassCons}. Since during the~evolution the~matter was scattered by the~gravitational potential barrier when the~collapsing shell approached its gravitational radius, the~mass conservation law was not satisfied in~the~entire domain of~integration. The~effect of~the~outgoing mass flux was negligible except the~vicinity of~the~event horizon. The~deviation from mass constancy increased with advanced time, as~the~event horizon was approached. The~maximal percentage deviations from mass conservation for~the~initial profiles P1 to~P9 are presented in~Table~\ref{tab:MassCons}. In~the~majority of~cases they were smaller than $1.0\%$. The~most significant deviation, exceeding $16\%$, was observed for~the~profile~P5. Table~\ref{tab:MassCons} also presents the~values of~$u$ corresponding to~the~event horizon locations and~to~the~percentage deviations not exceeding $5\%$. The~analysis of~mass conservation in~spacetime led to~the~conclusion that the~behavior of~matter investigated numerically for~all initial profiles was correct within the~domain of~integration. Only the~results obtained for~the~profile~P5 should be treated with reserve, but they were also not excluded from further analysis by the~performed code accuracy checks.

\begin{table*}
\caption{\label{tab:MassCons}The~maximal percentage deviations from mass conservation, DEV, the~values of~retarded time corresponding to~the~event horizon location, $u_{EH}$, and~to~the~percentage deviations from mass conservation not exceeding $5\%$, $u_5$, for~the~investigated initial profiles.}
\begin{ruledtabular}
\begin{tabular}{c|c|c|c|c|c|c|c|c|c}
Profile & P1 & P2 & P3 & P4 & P5 & P6 & P7 & P8 & P9 \\
\hline
DEV [$\%$] & $0.65$ & $0.7$ & $7.85$ & $0.62$ & $16.27$ & $0.97$ & $0.21$ & $2.53$ & $0.13$ \\
\hline
$u_{EH}$ & $1.0609$ & $0.2154$ & $0.3421$ & $0.4697$ & $0.4954$ & $1.2729$ & $0.2858$ & $1.2367$ & $0.9331$ \\
\hline
$u_{5}$ & $-$ & $-$ & $0.2631$ & $-$ & $0.1309$ & $-$ & $-$ & $-$ & $-$
\end{tabular}
\end{ruledtabular}
\end{table*}

During the~computations the~fulfillment of~equalities $f_{,v}=g_{,u}$ and~$x_{,v}=y_{,u}$, which stem from the~substitutions~(\ref{eqn:substitution}), was monitored for~all the~examined profiles. The~checks were performed in~both null directions along $v=v_f$ and~constant~$u$, whose values for~the~particular profiles were gathered in~the~caption of~Fig.~\ref{fig:Conv1}. The~necessary derivatives were calculated with the~use of~the~three-point regressive derivative method. The~relations $f_{,v}=g_{,u}$ and~$x_{,v}=y_{,u}$ were satisfied with an~accuracy of~$5.8\%$ and~$2.5\%$ along the~constant~$u$ hypersurfaces. The~accuracy at~$v=const.$ was equal to~$2.5\%$ and~$0.1\%$, respectively. The~discrepancies increased as~$r\to 0$ at~$u=const.$ and~as~$u\to u_{EH}$ at~constant~$v$. The~obtained accuracies of~the~constraints confirmed the~correctness of~the code.

\section{Results}
\label{sec:Results}

The~dynamical spacetime resulting from the~scalar field gravitational collapse is either non-singular or it~contains a~black hole of~a~Schwarzschild-type, i.e., a~spacelike singularity situated along $r=0$ surrounded by one horizon. The~final outcome of~the~process depends on~the~value of~the~amplitude $\amp$, which determines the~strength of~a~self-interaction of~the~scalar field. Since we were interested in~the~behavior of~the~scalar field in~regions of~high curvature, i.e.,~nearby the~singularity, we focused on~amplitudes leading to~the~formation of~a~black hole in~spacetime. They were selected for~every family of~initial profiles separately and~their values are given in~the~caption of~Fig.~\ref{fig:InitialProfiles}.

\begin{figure*}
\includegraphics[width=0.315\textwidth]{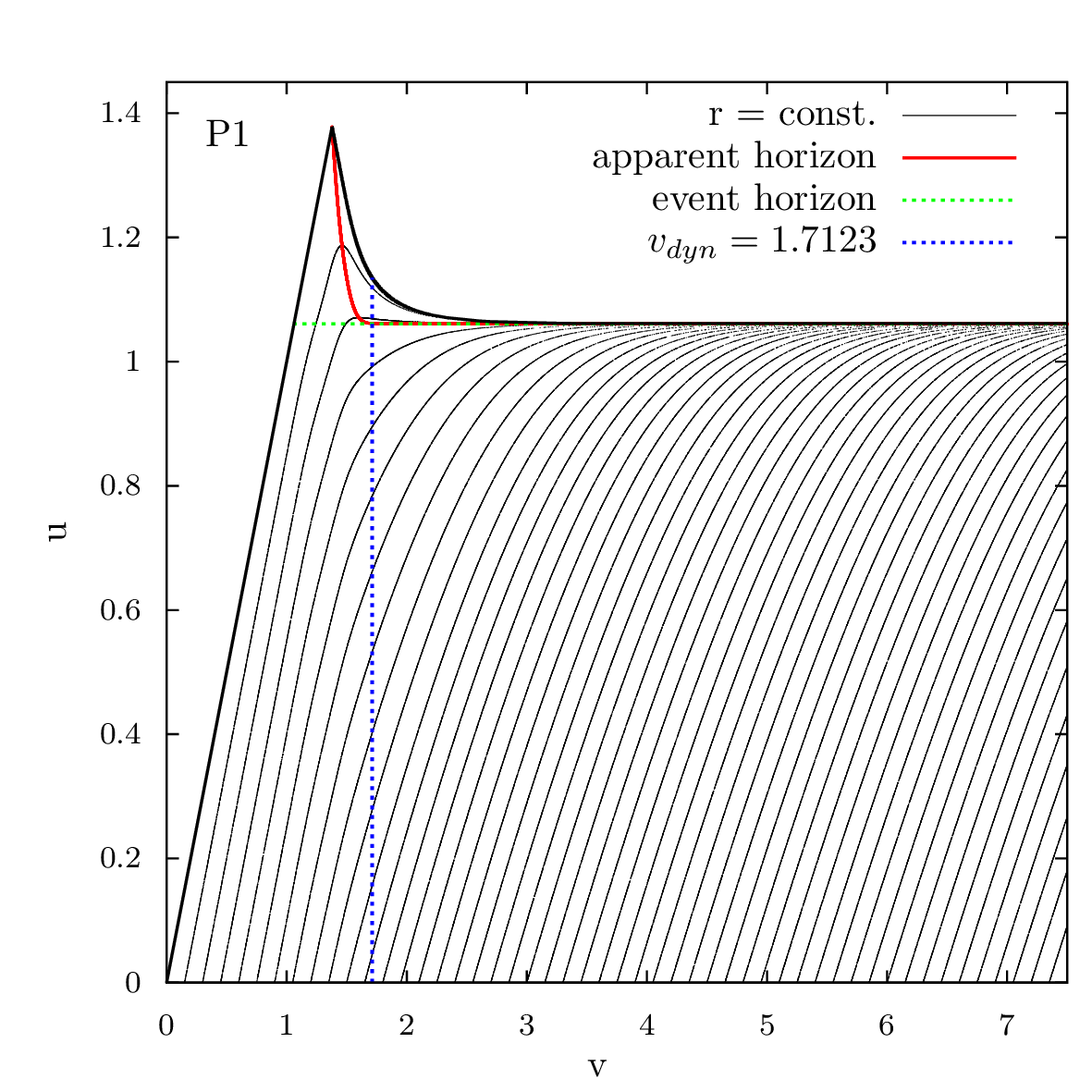}
\includegraphics[width=0.315\textwidth]{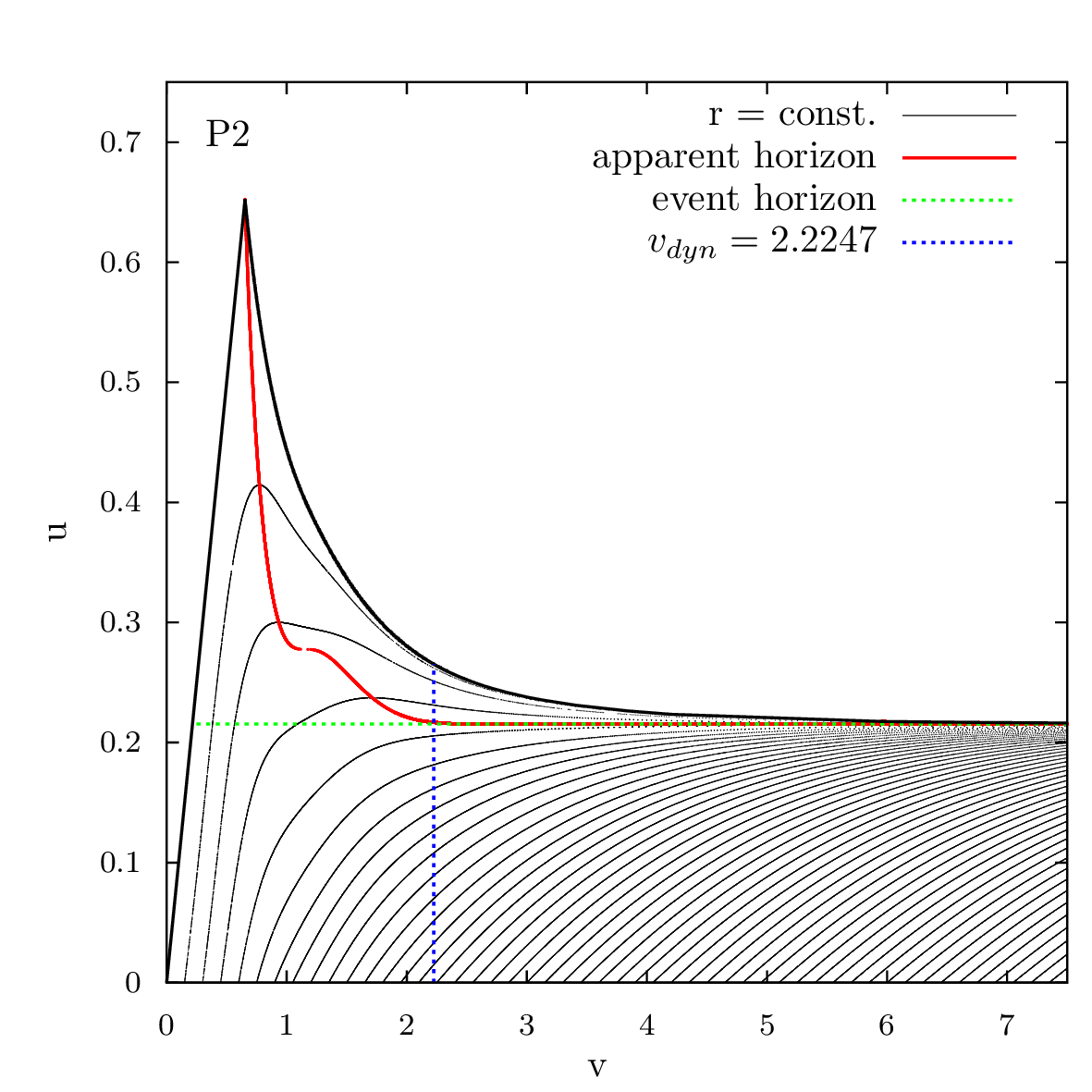}
\includegraphics[width=0.315\textwidth]{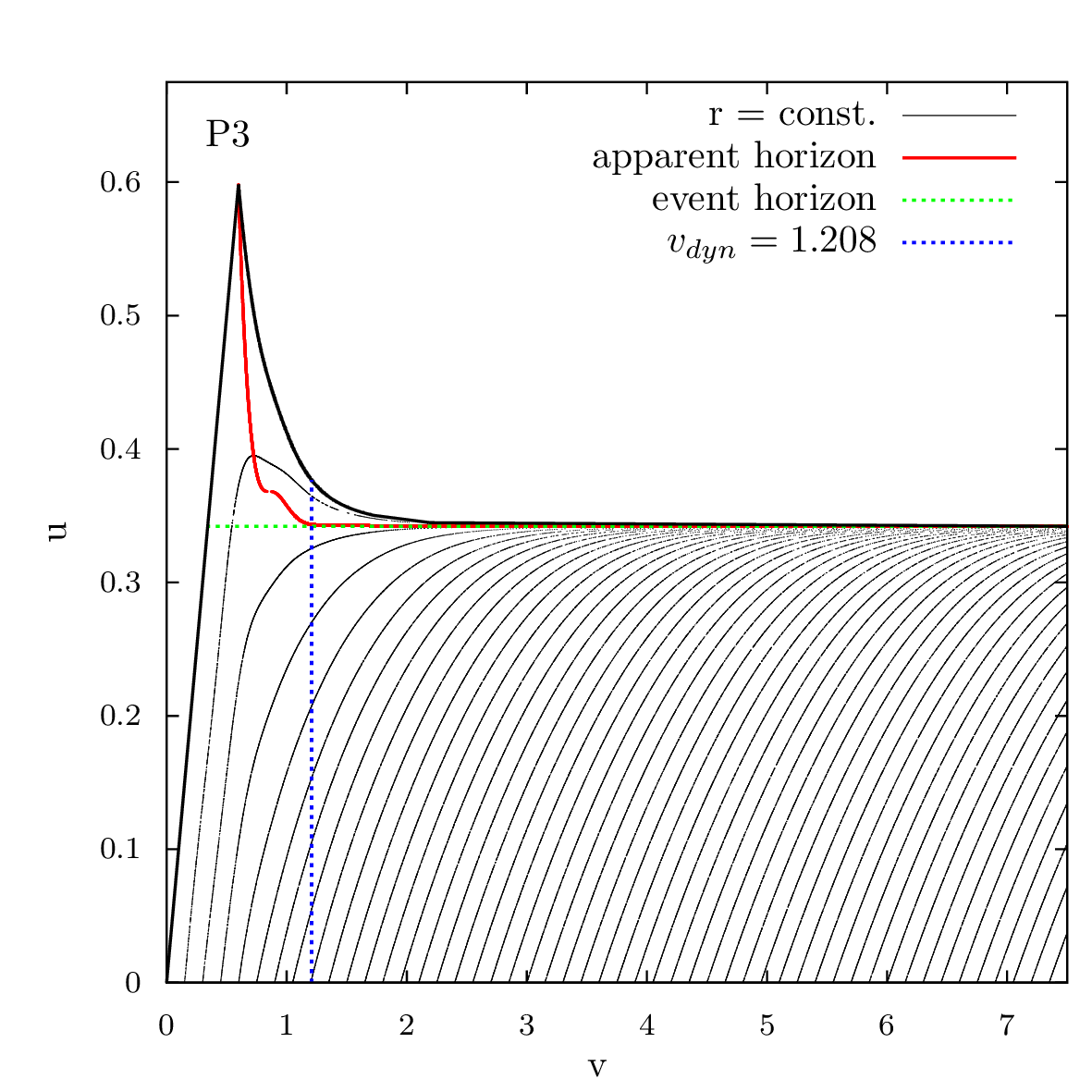}
\includegraphics[width=0.315\textwidth]{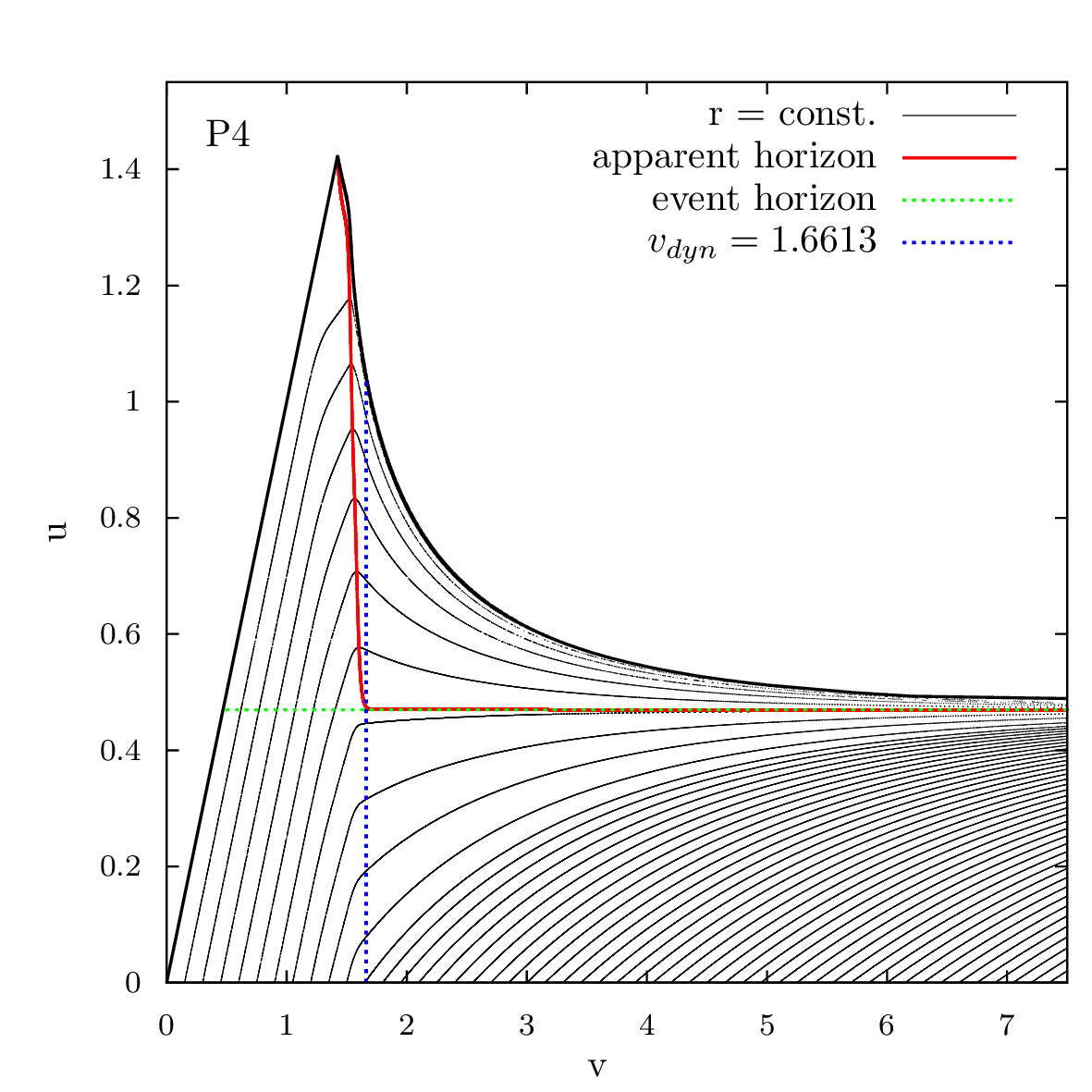}
\includegraphics[width=0.315\textwidth]{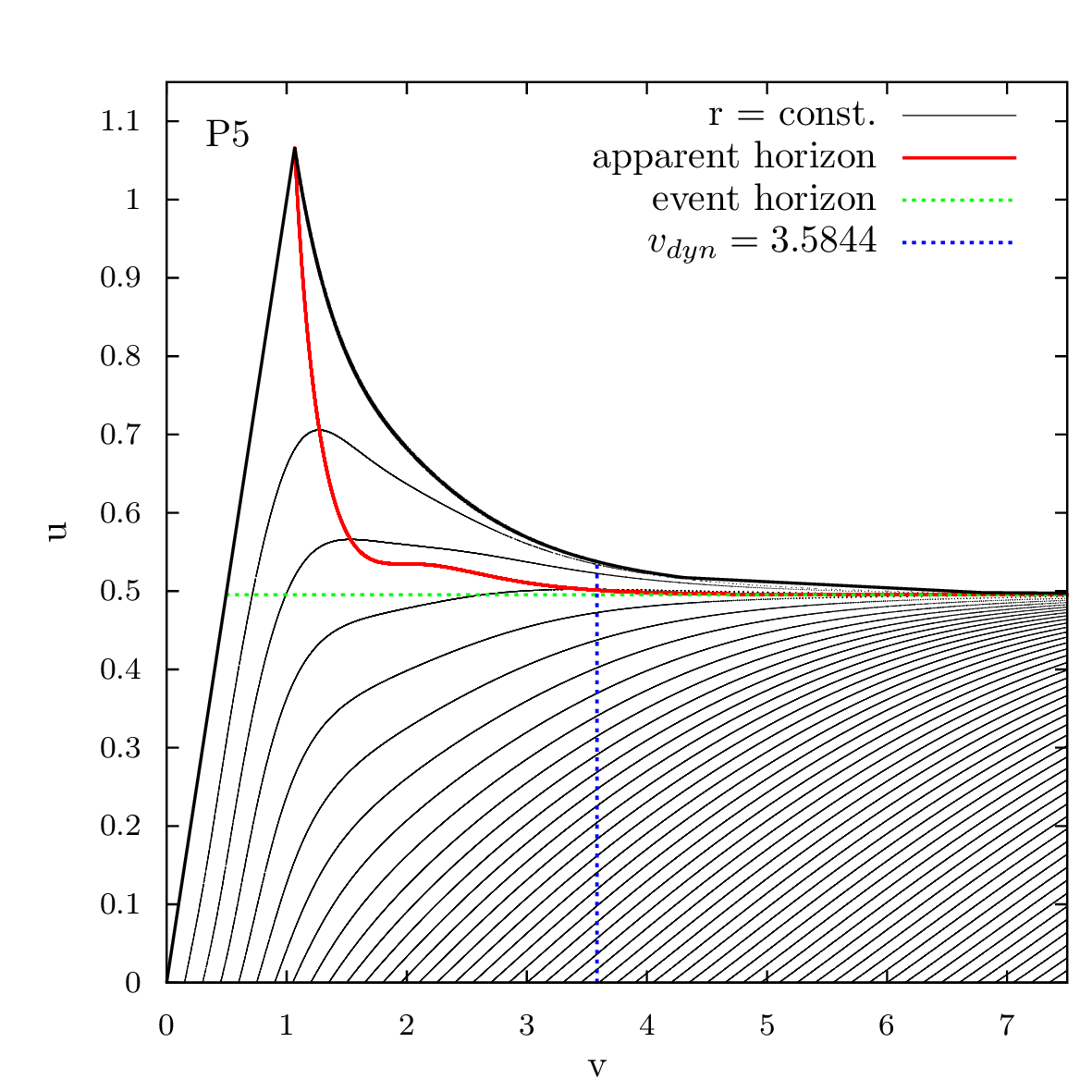}
\includegraphics[width=0.315\textwidth]{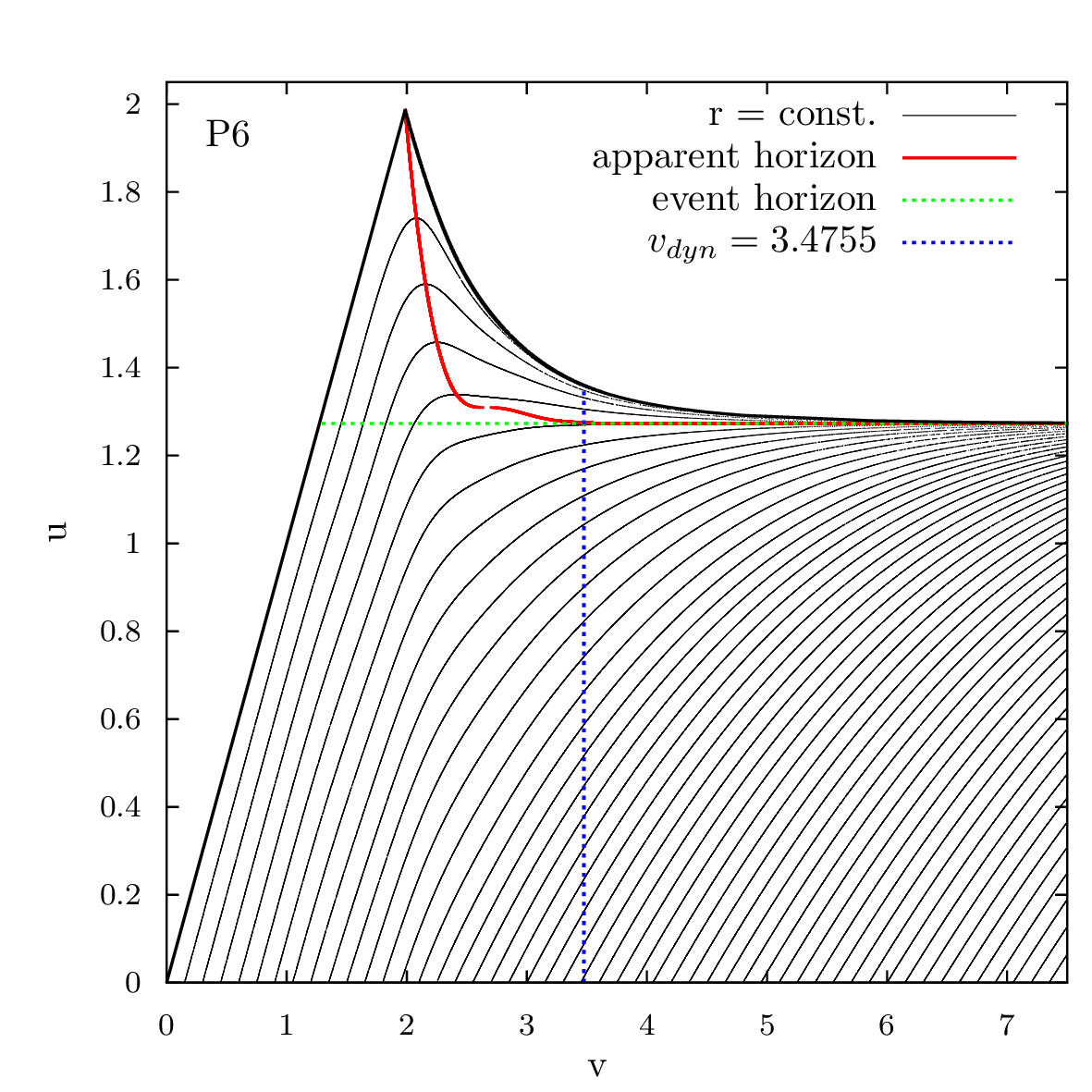}
\includegraphics[width=0.315\textwidth]{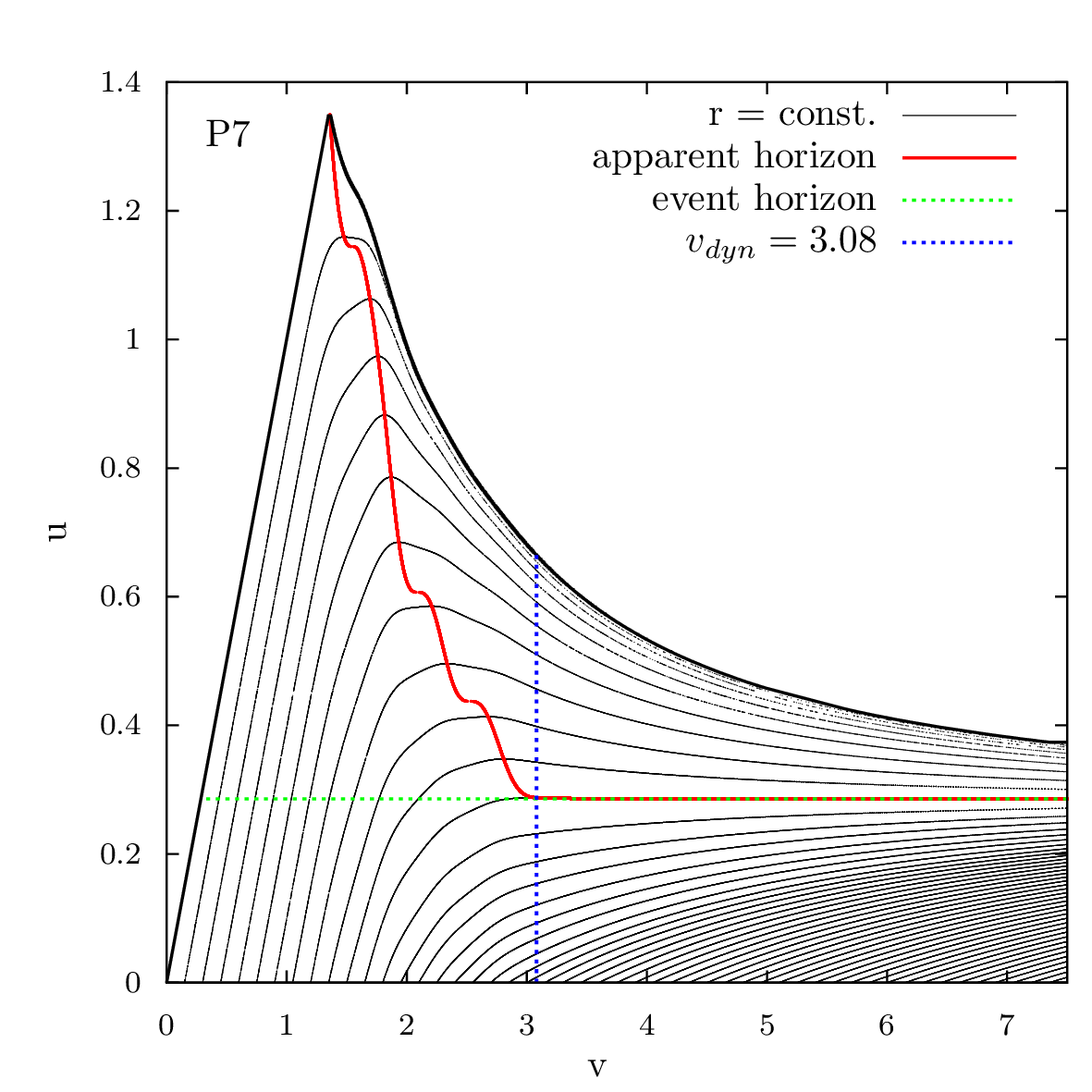}
\includegraphics[width=0.315\textwidth]{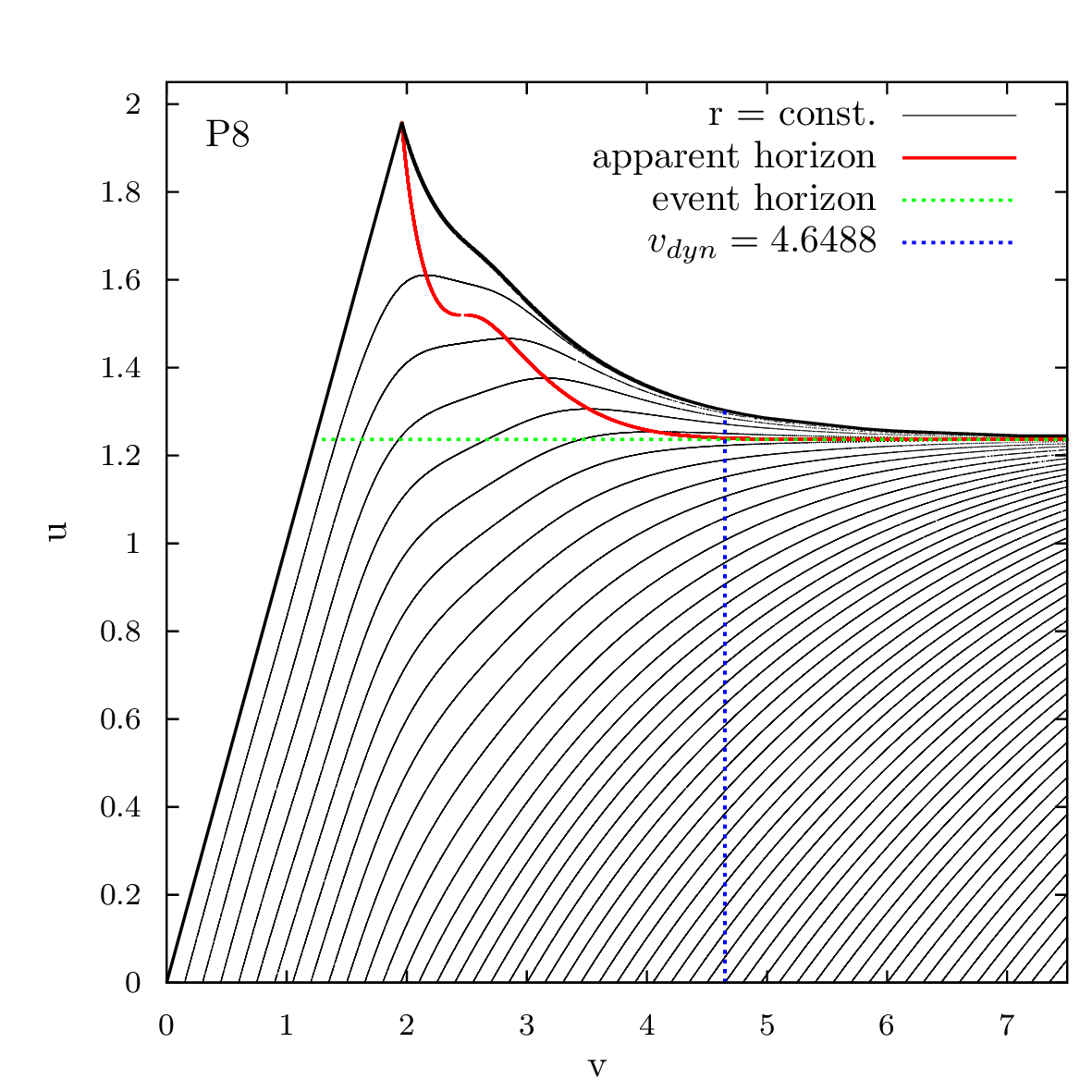}
\includegraphics[width=0.315\textwidth]{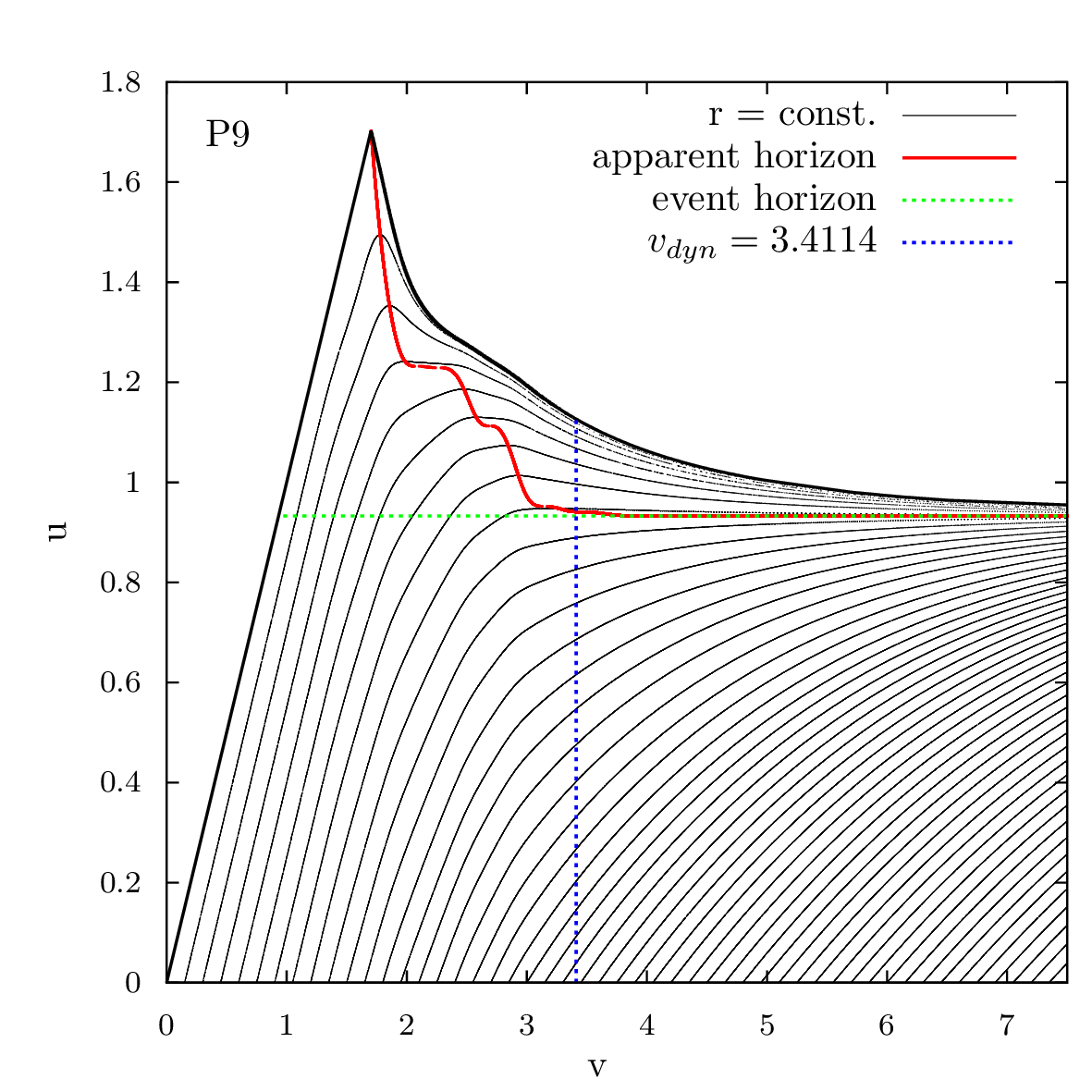}
\caption{(Color online) Penrose diagrams of~dynamical Schwarzschild spacetimes formed during the~scalar field gravitational collapse with initial profiles P1 to~P9. The~borders between dynamical and~non-dynamical regions are situated along null hypersurfaces denoted as~$v_{dyn}$.}
\label{fig:PenroseDiagrams}
\end{figure*}

Penrose diagrams of~the~obtained dynamical spacetimes, which present $r=const.$ lines in~the~$\left(vu\right)$-plane with $\theta$ and~$\phi$ suppressed, are shown in~Fig.~\ref{fig:PenroseDiagrams}. The~central singularity is situated along the~non-linear part of~the~outermost $r=const.$ line, which was bolded on~the~diagrams. It~is spacelike, because the~derivative $\frac{du}{dv}\big|_{r=0}$ is negative along the~line. The~singularity is surrounded by a~single apparent horizon, marked red on~the~diagrams, which is situated along a~hypersurface, which becomes null in~the~region where $v\rightarrow\infty$. As~an~apparent horizon coincides with an~event horizon in~static spacetimes, it~is postulated that its null character at~infinity indicates the~location of~the~event horizon in~spacetime. Event horizons of~the~emerging spacetimes were presented as~green dashed null lines on~the~Penrose diagrams.

The~location of~an apparent horizon in~the~particular spacetime allows us to~divide the~neighborhood of~the~central singularity into two parts. The~first one corresponds to~the~range of~the~$v$-coordinate, within which the~apparent horizon changes its position along $u$. This region will be referred to~as dynamical, because the~apparent horizon is situated inside the~event horizon, which is a~characteristic property of~dynamical spacetimes~\cite{HawkingEllis}. In~this area the~actual dynamics of~the~collapse takes place, before the~spacetime settles down as~$v\rightarrow\infty$. The~remaining part of~the~vicinity of~the~singularity is related to~the~apparent horizon being situated along a~null hypersurface. This region will be dubbed as~non-dynamical. The~null hypersurfaces of~constant $v$ which are borders between the~dynamical and~non-dynamical regions for~spacetimes obtained with the~use of~profiles P1 to~P9 are pictured on~the~plots in~Fig.~\ref{fig:PenroseDiagrams} as~blue dashed lines. The~respective values of~advanced time, $v_{dyn}$, are presented on~the~adequate diagrams.

The~values of~$v_{dyn}$ were calculated on~the~basis of~examining the~changes of~$u$ and~$v$-coordinates along the~apparent horizon. The~changes of~retarded time ($\delta u$) are constant due to~the~construction of~the~numerical grid (see Section~\ref{sec:NumComp}). The~differences in~advanced time ($\delta v$) on~adjacent $u=const.$ hypersurfaces vary from $\delta u$ to~the~values of~an order of~$10^4\delta u$. The~values of~$\delta v$ close to~$\delta u$ are characteristic for~the~dynamical region, while the~growth of~$\delta v$ means that the~apparent horizon settles along $u=const.$ indicating the~non-dynamical area. The~selected value of~$v_{dyn}$ was the~biggest value of~advanced time corresponding to~$\delta v\geqslant 10^2\delta u$.

\begin{figure*}
\includegraphics[width=0.315\textwidth]{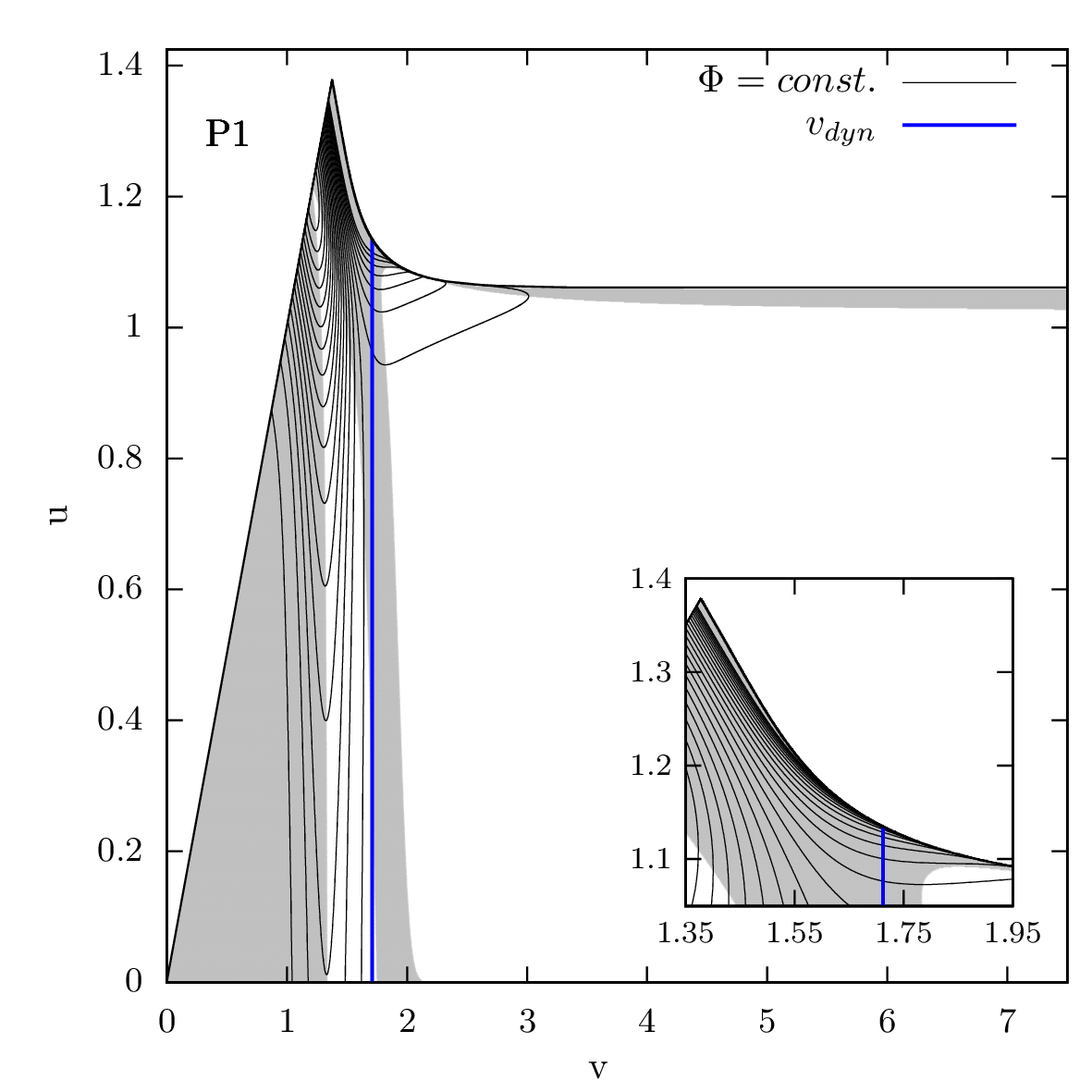}
\includegraphics[width=0.315\textwidth]{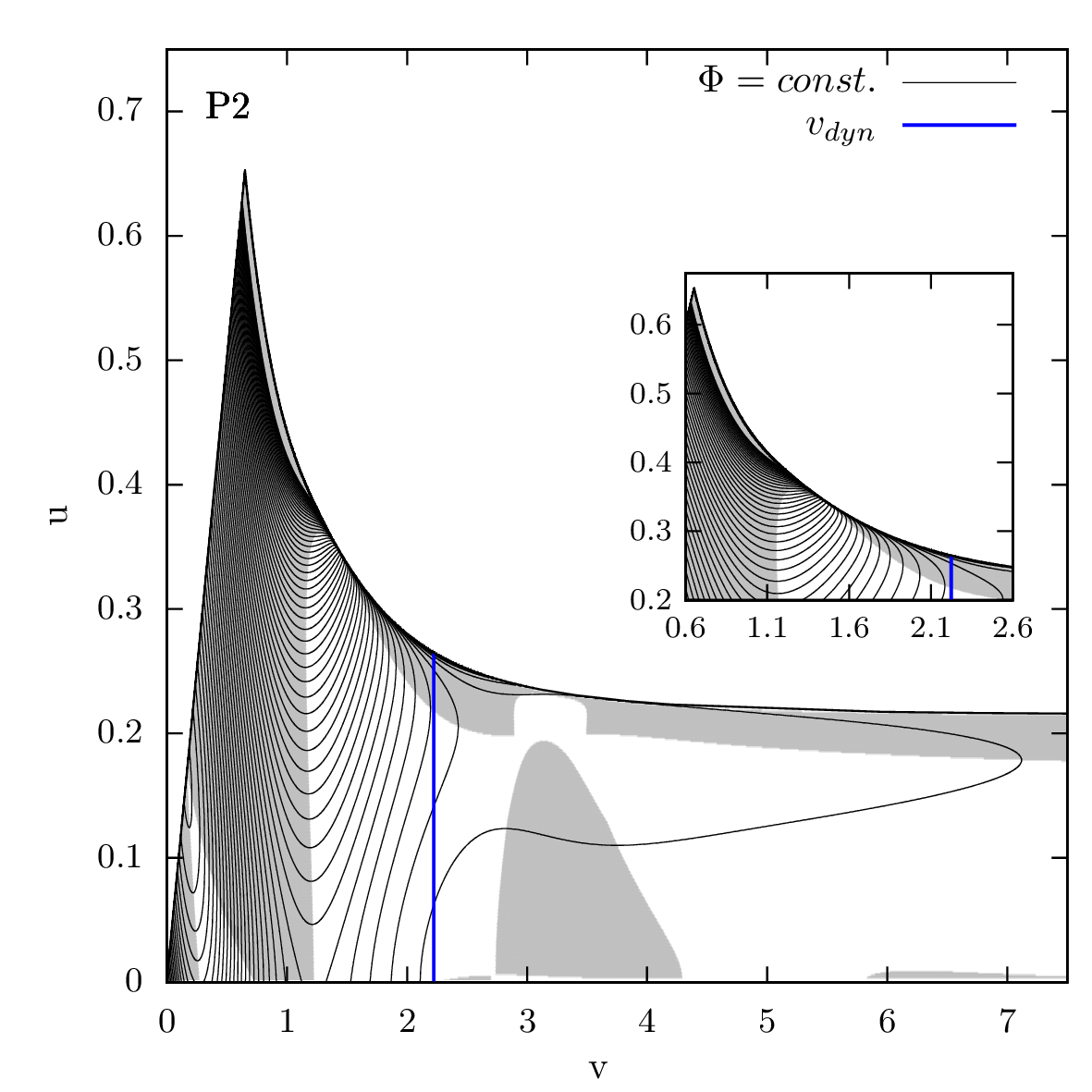}
\includegraphics[width=0.315\textwidth]{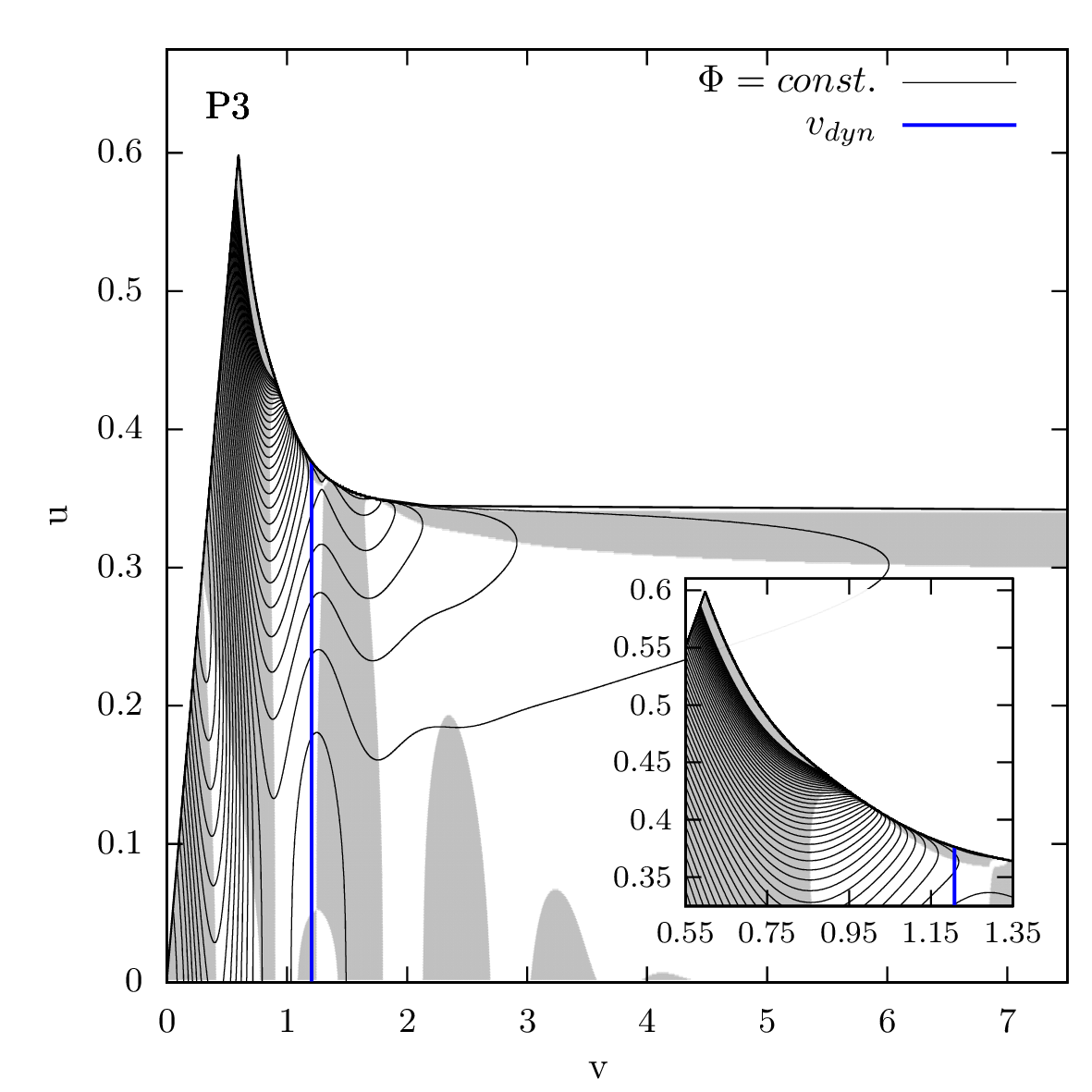}
\includegraphics[width=0.315\textwidth]{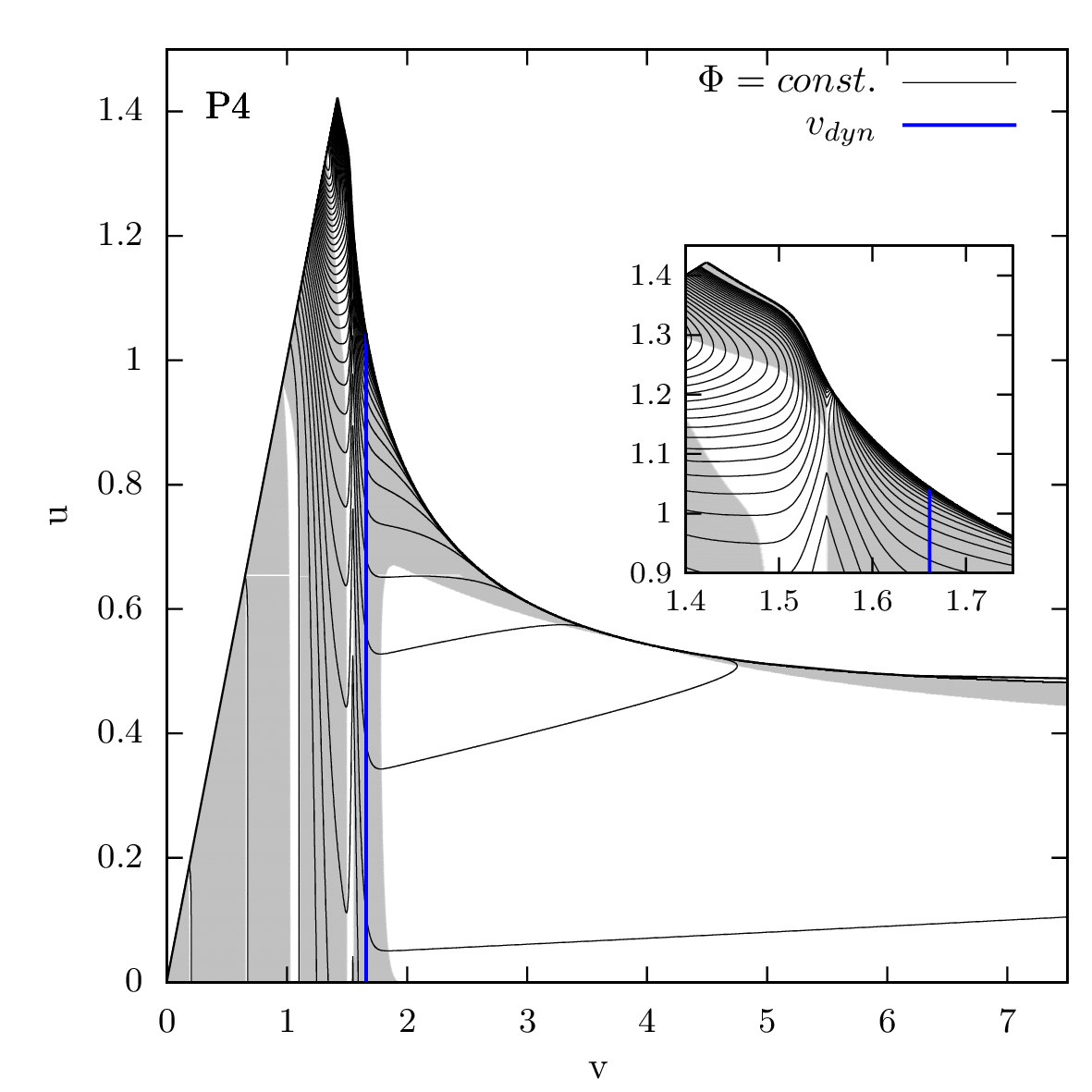}
\includegraphics[width=0.315\textwidth]{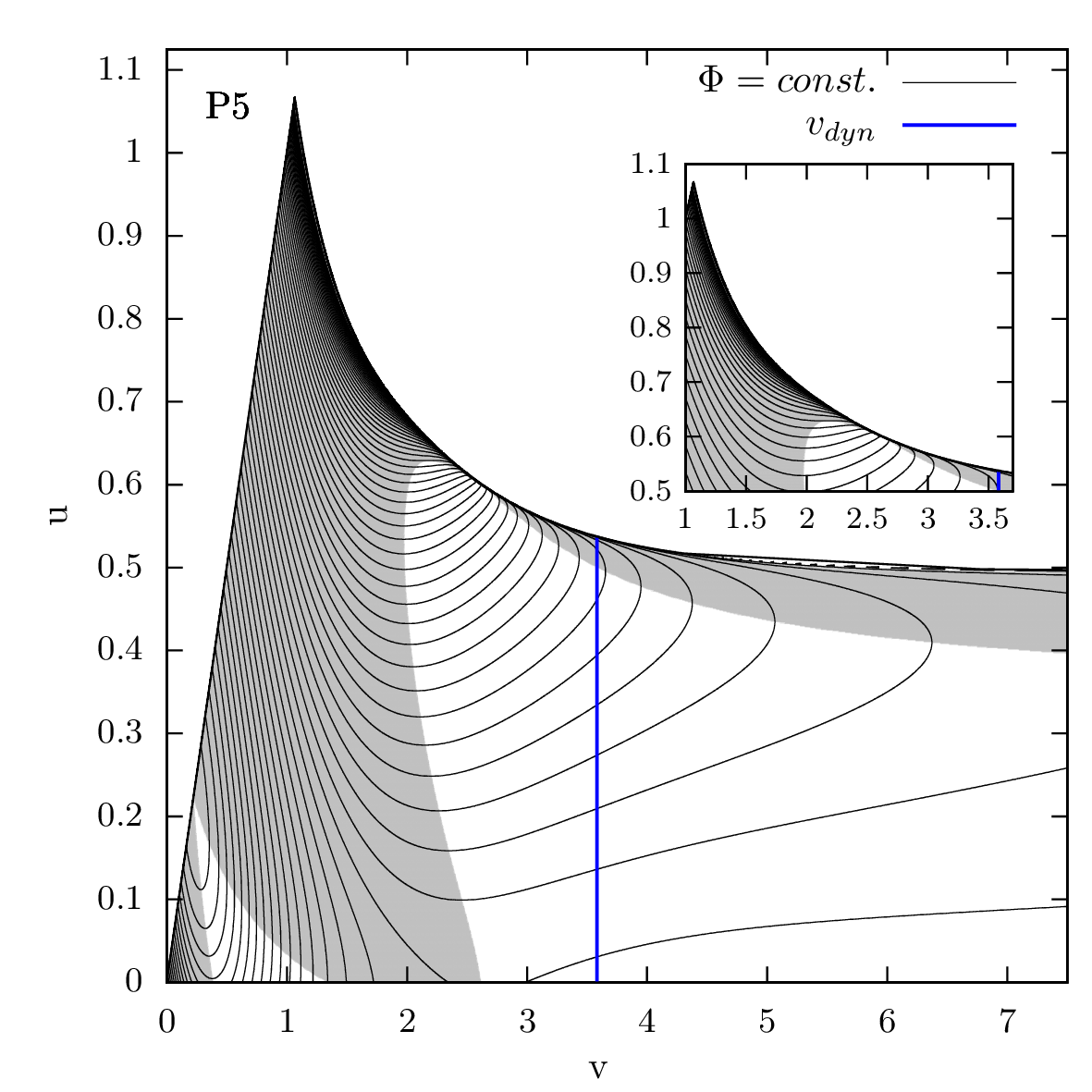}
\includegraphics[width=0.315\textwidth]{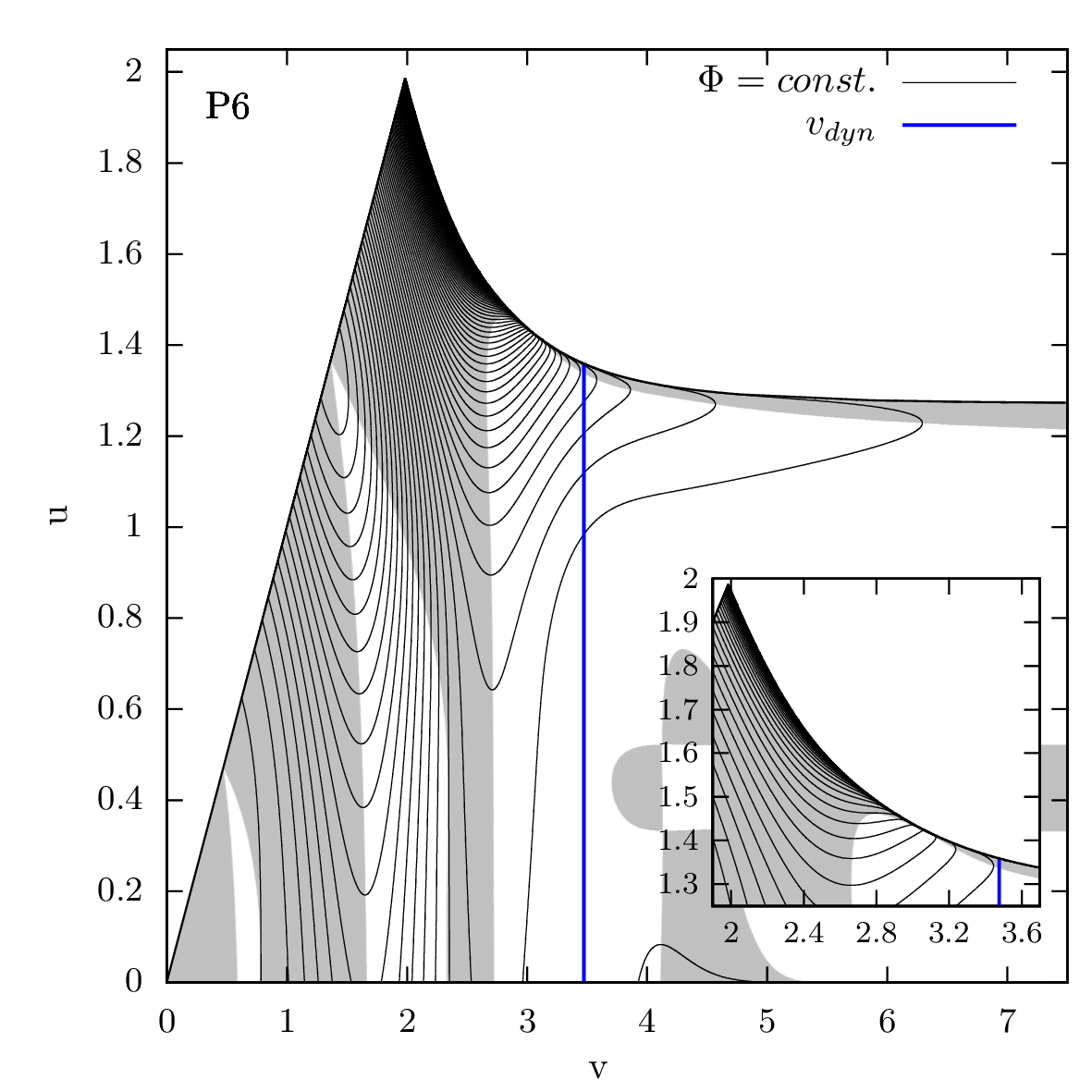}
\includegraphics[width=0.315\textwidth]{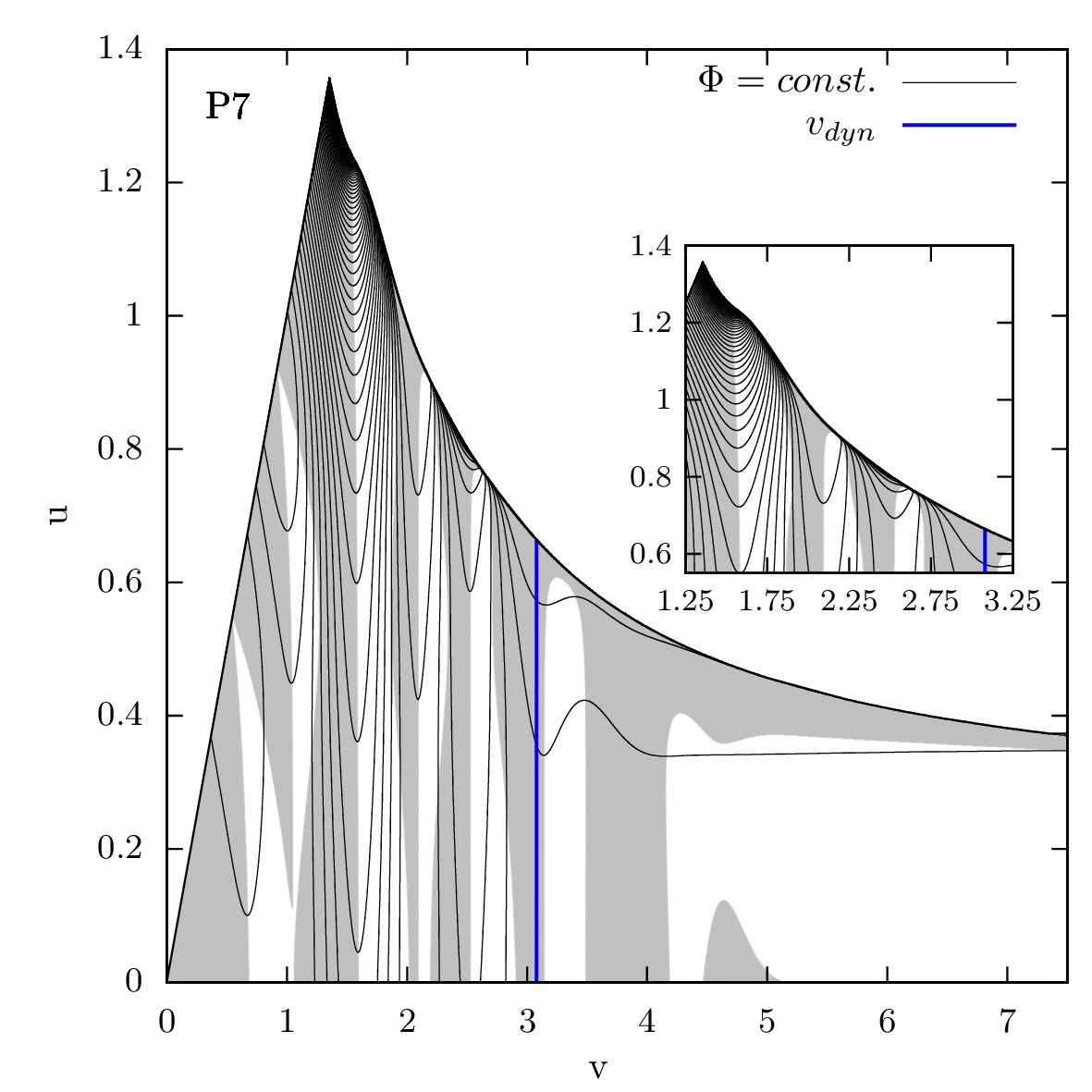}
\includegraphics[width=0.315\textwidth]{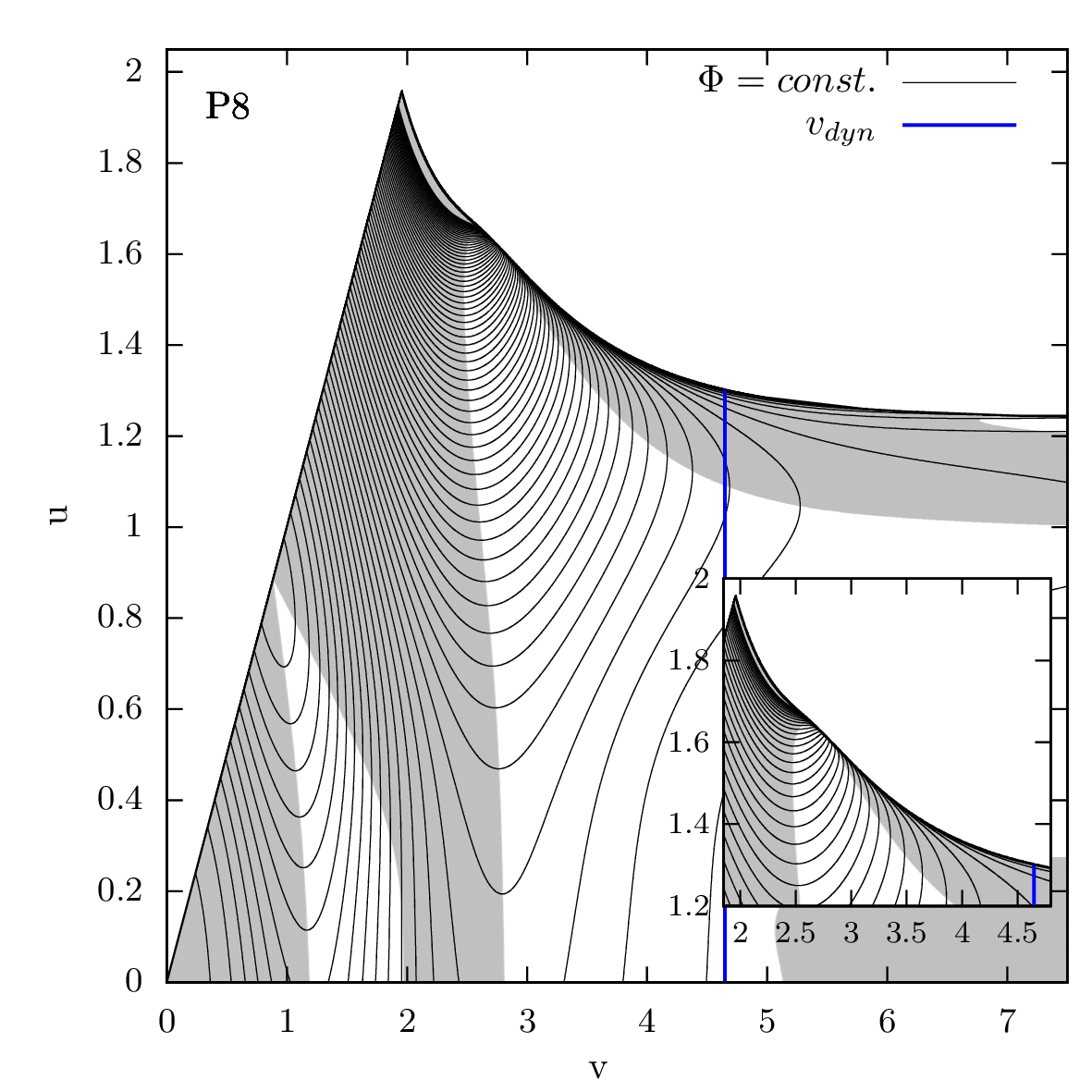}
\includegraphics[width=0.315\textwidth]{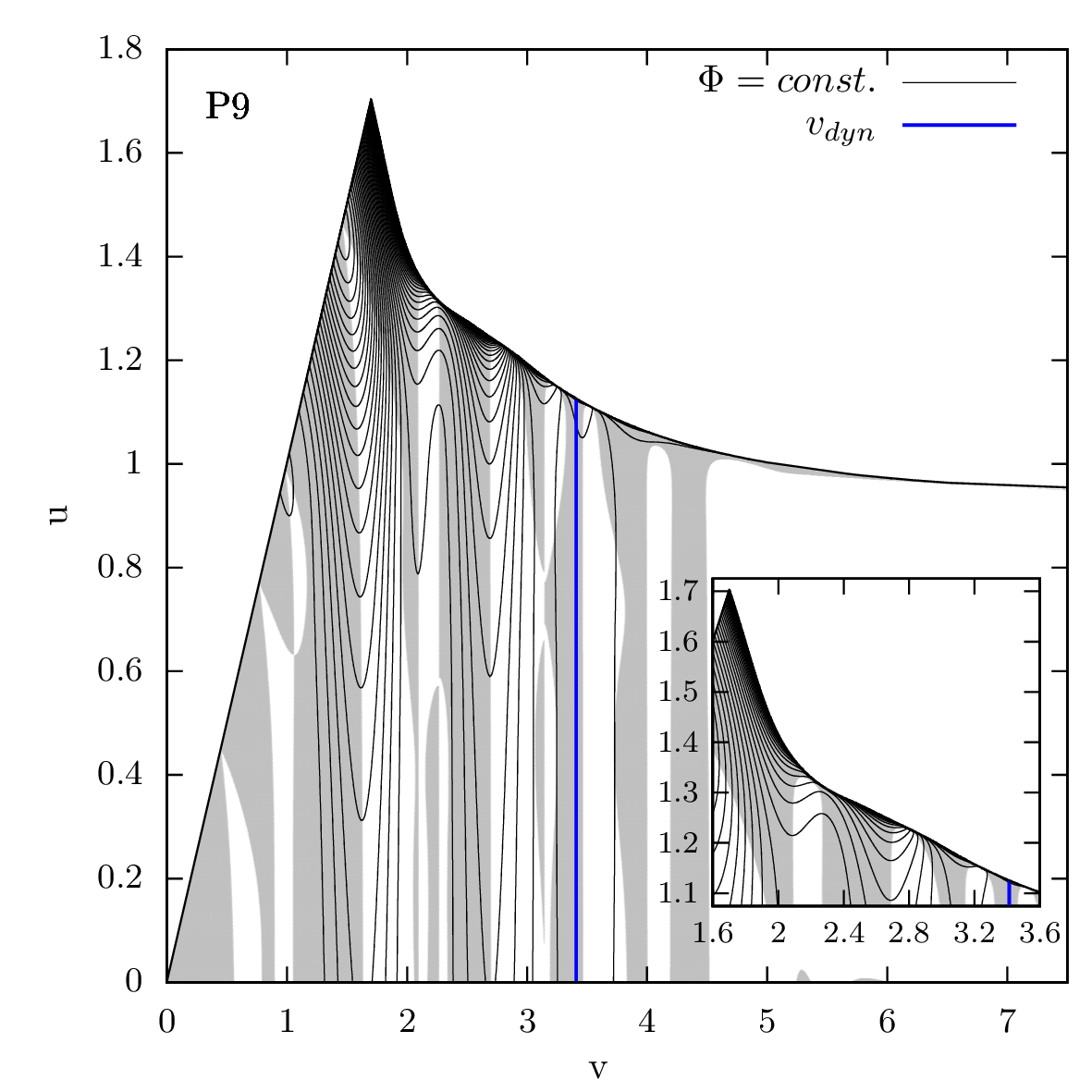}
\caption{(Color online) Lines of~constant $\Phi$ in~spacetimes developed during the~gravitational collapse of~a~scalar field with initial profiles~P1 to~P9. Gray areas indicate spacetime regions, in~which the~hypersurfaces $\Phi=const.$ are spacelike. The~dynamical regions in~the~vicinities of~central singularities were magnified.}
\label{fig:ConstPhiSl}
\end{figure*}

The~condition which has to~be fulfilled in~order to~treat the~dynamical scalar field as~a~time variable is that the~lines of~constant values of~the~field are spacelike. The~trajectories of~$\Phi=const.$ lines in~spacetimes emerging from the~collapse initiated with the~use of~the~profiles P1 to~P9 are presented in~Fig.~\ref{fig:ConstPhiSl}. The~spacetime regions, in~which the~lines are spacelike were marked gray on~the~plots and~the~dynamical areas nearby central singularities were magnified. The~character of~the~particular constant $\Phi$ line was determined on~the~basis of~the~sign of~the~derivative $\frac{du}{dv}\big|_{\Phi=const.}$ calculated along~it. The~areas of~spacelike and~timelike hypersurfaces $\Phi=const.$ were characterized by positive and~negative values of~the~derivatives, respectively.

It~may be inferred from the~diagram for~the~profile~P1 that in~the~vicinity of~the~singularity the~lines of~constant $\Phi$ are spacelike, except for~a~limited area, in~which the~lines are timelike or null and~which seems to~have one boundary point situated at~the~singularity. The~area is situated outside of~the~dynamical part of~the~vicinity of~the~singularity. Hence, nearby the~central singularity the~scalar field can be treated as~a~time variable.

The~interpretation of~the~results obtained for~the~remaining initial profiles is analogous, having in~mind two disparities. Firstly, the~amount of~the~above-mentioned areas, whose boundaries have single common points with the~singularity, depend on~the~family of~the~initial profile and~these areas can be situated within the~dynamical regions nearby the~singularity (e.g.,~profiles P3, P7 and~P9). Secondly, there can exist certain regions situated in~non-dynamical parts of~the~neighborhoods of~the~singularities, in~which the~$\Phi=const.$ lines are non-spacelike (e.g.,~profiles P4, P8 and~P9). Despite the~outlined differences, the~results obtained for~all the~profiles enhance the~comprehensiveness of~the~conclusion that a~scalar field can be treated as~a~time variable safely, at~least in~the~dynamical part of~the~vicinity of~the~singularity of~the~evolving spacetime.

Another requirement, which is crucial for~treating the~scalar field as~a~time variable is its monotonicity in~the~regions of~interest. An~inspection of~the~course of~lines of~constant $\Phi$ shown on~the~plots in~Fig.~\ref{fig:ConstPhiSl} allows us to~draw a~conclusion that in~the~regions, where the~lines are spacelike, the~changes are monotonic along constant $u$ and~$v$. Since the~time coordinate is defined as~$t=\frac{u+v}{2}$ in~double null coordinates and~the~$t$-axis forms an~angle of~45$^\circ$ with both $u$ and~$v$ axes, the~above conclusion can be extended to~the~coordinate~$t$. The~non-monotonic behavior of~$\Phi=const.$ lines arises only when areas of~spacelike and~timelike or null characters are considered, which is excluded regarding the~aforementioned demand of~the~spacelike nature of~constancy hypersurfaces of~$\Phi$.

An~important aspect of~the~conducted analysis, which stems from the~fact that the~numerical methods were employed, is whether the~singularity was approached closely enough in~order to~draw conclusions about the~behavior of~the~evolving field in~its close proximity. The~estimation of~the~singularity closeness which was achieved during numerical computations is presented in~Table~\ref{tab:Closeness}. It~was made along the~whole central singularity on~the~basis of~the~relation between the~value of~the~radial function $r$ which refers to~the~location of~the~event horizon in~spacetime, $r_{EH}$, and~its lowest value obtained in~the~course of~simulations, $r_{min}$. It turned out that the~minimal values of~the~radial function are at~least six orders of~magnitude smaller in~comparison to~the~ones which refer to~the~event horizon location. This defines the~relative degree of~numerical closeness to~the~singularity obtained during the~evolution of~the~scalar field.

\begin{table*}
\caption{\label{tab:Closeness}Computational closeness to~the~singularity. $r_{EH}$ denotes the~value of~the~radial function related to~the~event horizon location in~spacetime, while $r_{min}$ is the~minimum value of~$r$, which was achieved numerically in~the~vicinity of~the~central singularity.}
\begin{ruledtabular}
\begin{tabular}{c|c|c|c|c|c|c|c|c|c}
Profile & P1 & P2 & P3 & P4 & P5 & P6 & P7 & P8 & P9 \\
\hline
$r_{EH}$ & $0.16003$ & $0.23711$ & $0.12424$ & $0.51198$ & $0.20897$ & $0.36999$ & $0.78311$ & $0.41127$ & $0.62098$ \\
\hline
$r_{min}\cdot 10^7$ & $11.23$ & $2.60$ & $1.88$ & $0.48$ & $0.20$ & $18.72$ & $1.74$ & $0.36$ & $52.65$ \\
\hline
$\frac{r_{min}}{r_{EH}}\cdot 10^7$ & $70.17$ & $10.97$ & $15.13$ & $0.94$ & $0.93$ & $50.60$ & $2.22$ & $0.88$ & $84.79$
\end{tabular}
\end{ruledtabular}
\end{table*}

It~should be stressed that the~presented analysis is valid for~the~investigated families of~profiles presented in~Fig.~\ref{fig:InitialProfiles}, which means that it~is independent of~the~selected values of~initial amplitudes~$\amp$. Due to~the~widely accepted property of~the~scalar field dynamical collapse, which is its independence of~the~type of~the~initial profile~\cite{Choptuik1993}, the~conclusions can be also extended to~other profiles having properties described at~the~end of~Section~\ref{sec:TheorFrame}.

\section{Conclusions}
\label{sec:Conclusions}

The~possibility of~investigating dynamics of~quantum gravitational systems with respect to~one of~their matter components has been successfully exploited in~various researches within quantum gravity. In~our paper we made an~attempt to~justify using a~scalar field as~an~intrinsic 'clock' during investigations of~dynamical systems. For~this purpose, we analyzed a~dynamical evolution of~a~self-interacting scalar field within the~framework of~general relativity, for~a~collection of~nine different initial configurations of~the~field.

As~a~result of~each evolution a~Schwarzschild spacetime was obtained. It contained a~spacelike singularity surrounded by an~apparent horizon, which for~all investigated initial conditions was spacelike in~the~dynamical part of~the~spacetime. This is an~interesting observation when associated with the~notion of~dynamical horizons, which are quasi-local spacelike horizons used to~investigate numerous diversified aspects of~physics of~evolving black holes~\cite{AshtekarKrishnan2004}.

The~interpretation of~the~results was concentrated on~the~issue, whether the~evolving scalar field fulfills the~necessary requirements to~be regarded as~a~time variable. Our~studies led to~the~following conclusions.

\begin{enumerate}[leftmargin=*]

 \item The~lines of~constant scalar field are spacelike in~the~dynamical part of~the~vicinity of~the~central spacelike singularity formed during the~dynamical gravitational collapse for~all the~investigated families of~initial profiles. For families F4, F8 and~F9 there exist regions situated in~non-dynamical areas, in~which the~lines are timelike or null.
 
 \item The~scalar field values change monotonically with $u$, $v$ and~$t$ coordinates in~the~regions, in~which the~hypersurfaces of~constant $\Phi$ are spacelike.
 
 \item The~distance from the~central singularity achieved numerically was estimated in~relation to~the~location of~the~event horizon in~spacetime. Although numerical computations did not allow us to~trace the~scalar field evolution precisely up to~the~central singularity, the~relative distance from it~was at~least six orders of~magnitude smaller than the~event horizon location, which means that it~was highly satisfactory in~the~context of~investigating the~behavior of~the~field in~its vicinity.
 
\end{enumerate}

The~above findings revealed that the~scalar field can be treated as~a~time variable in~dynamical parts of~evolving spacetimes for~certain, while the~non-dynamical parts should be treated cautiously. It~should be stressed that the~presented outcomes are valid for~the~massless scalar field coupled to~gravity only. In~the~case of~more complicated gravitational systems, using matter which is coupled not only to~gravity, but~also to~additional matter fields, as~a~time measurer requires separate investigations, probably adapted to~the~specific problem. Some indications that matter-geometry systems, which include non-minimally coupled Brans-Dicke field and~a~real or~complex scalar field can admit intrinsic 'clocks' can be found in~\cite{HwangYeom2010,HansenYeom2014}.

\begin{acknowledgments}
This work was partially supported by the~Polish National Science Center Grant No.~2011/02/A/ST2/00300.
\end{acknowledgments}

\bibliography{TimeVarConstPHI.bib}

\end{document}